\documentclass[10pt]{iopart}

\eqnobysec
\def \BEA { \begin{eqnarray}}
\def \EEA {\end{eqnarray}}
\def \BE {\begin{equation}}
\def \EE {\end{equation}}

\def \nn {\nonumber}

\def \kd #1 {\delta_{#1}}

\def \eq5d {\stackrel{{\mbox{\tiny (n=5)}}}{=}}

\def \bl {\mbox{\boldmath{$\ell$}}}
\def \hbl {\mbox{\boldmath{$\hat \ell$}}}
\def \bn {\mbox{\boldmath{$n$}}}

\def \hbn {\mbox{\boldmath{$\hat n$}}}
\def \hbm #1 {\mbox{\boldmath{$\hat m^{(#1)}$}}}
\def \bm {\mbox{\boldmath{$m$}}}

\def \vbm #1 {\mbox{\boldmath{$m^{({#1})}$}}}

\def \WDS #1 {\mbox{$\Phi_{#1}^{S}$}}
\def \WDA #1 {\mbox{$\Phi_{#1}^{A}$}}
\def \WD #1 {\mbox{$\Phi_{#1}$}}

\def \Mi {\stackrel{i}{M}}
\def \Mj {\stackrel{j}{M}}
\def \Mk {\stackrel{k}{M}}

\def \Ms {\stackrel{s}{M}}
\def \Me {\stackrel{e}{M}}
\def \Ml {\stackrel{l}{M}}
\def \Mp {\stackrel{p}{M}}

\newcommand{\hM}[3] {{\stackrel{#1}{\hat{M}}}_{{#2}{#3}}}

\def \mi {\stackrel{i}{m}}
\def \mj {\stackrel{j}{m}}
\def \mk {\stackrel{k}{m}}
\def \mr {\stackrel{r}{m}}
\def \ms {\stackrel{s}{m}}
\def \mp {\stackrel{p}{m}}
\def \mz {\stackrel{z}{m}}
\def \mq {\stackrel{q}{m}}
\def \mo {\stackrel{o}{m}}
\def \mD {\stackrel{2}{m}}
\def \mT {\stackrel{3}{m}}
\def \mC {\stackrel{4}{m}}

\def \mio #1 {\mi_{#1}\ ^{  \! \! \! \! 0}} 
\def \mjo #1 {\mj_{#1}\ ^{  \! \! \! \! 0}} 
\def \mko #1 {\mk_{#1}\ ^{  \! \! \! \! 0}} 
\def \mro #1 {\mr_{#1}\ ^{  \! \! \! \! 0}} 
\def \mso #1 {\ms_{#1}\ ^{  \! \! \! \! 0}} 
\def \mpo #1 {\mp_{#1}\ ^{  \! \! \! \! 0}} 
\def \mzo #1 {\mz_{#1}\ ^{  \! \! \! \! 0}} 
\def \mqo #1 {\mq_{#1}\ ^{  \! \! \! \! 0}} 
\def \moo #1 {\mo_{#1}\ ^{  \! \! \! \! 0}} 
\def \mDo #1 {\mD_{#1}\ ^{  \! \! \! \! 0}} 
\def \mTo #1 {\mT_{#1}\ ^{  \! \! \! \! 0}} 
\def \mCo #1 {\mC_{#1}\ ^{  \! \! \! \! 0}}

\newcommand{\be}{\begin{equation}}
\newcommand{\ee}{\end{equation}}
\newcommand{\beqn}{\begin{eqnarray}}
\newcommand{\eeqn}{\end{eqnarray}}

\newcommand{\ba}{\begin{array}}
\newcommand{\ea}{\end{array}}
\newcommand{\pp}{{\it pp\,}-}

\def \cO {{\cal O}}

\def \BEAH {\begin{eqnarray*}}
\def \EEAH {\end{eqnarray*}}
\def \BEA {\begin{eqnarray}}
\def \EEA {\end{eqnarray}}
\def \BDM {\begin{displaymath}}
\def \EDM {\end{displaymath}}
\def \Mi {\stackrel{i}{M}}
\def \Mj {\stackrel{j}{M}}
\def \Mk {\stackrel{k}{M}}

\def \Ms {\stackrel{s}{M}}

\def \pul {\textstyle{\frac{1}{2}}}

\def \T {\bigtriangleup  }

\begin{document}
\renewcommand{\thefootnote}{\arabic{footnote}}

\title[Vacuum spacetimes with a  non-twisting geodetic multiple WAND]{Newman-Penrose formalism in higher dimensions:  vacuum spacetimes with a non-twisting geodetic multiple Weyl aligned null direction}

\author{A. Pravdov\' a, V. Pravda}

\address{ Mathematical Institute, 
Academy of Sciences, \v Zitn\' a 25, 115 67 Prague 1, Czech Republic}
\eads{\mailto{pravdova@math.cas.cz}, \mailto{pravda@math.cas.cz} }

\begin{abstract}
Vacuum spacetimes admitting a non-twisting geodetic multiple Weyl aligned null direction (WAND) are analyzed in arbitrary dimension using recently developed higher-dimensional Newman-Penrose (NP) formalism. We determine dependence of the metric and of the Weyl tensor on the affine parameter $r$ along null geodesics generated by the WAND for type III and N spacetimes and for a special class of type II and D spacetimes, containing e.g. Schwarzschild-Tangherlini black holes and black strings and branes.

For types III and N, all metric components are at most quadratic polynomials in $r$ while for types II and D the $r$-dependence of the metric as well as of the Weyl tensor is determined by an integer $m$ corresponding to the rank of the expansion matrix $S_{ij}$.  It is shown that for non-vanishing expansion, all these spacetimes contain a curvature singularity.

As an illustrative example, a shearing expanding type N five-dimensional vacuum solution is also re-derived using higher-dimensional NP formalism. This solution can be, however,  identified with a direct product of a known four-dimensional type N metric with an extra dimension.
\end{abstract} 

%\pacs{?04.50.+h, 04.20.-q, 04.20.Cv}

\section{Introduction}

The null frame Newman-Penrose (NP) formalism \cite{NP,Stephanibook} is a very useful tool for constructing exact  solutions of the four-dimensional general relativity. Although the number of equations  is considerably larger than in the standard coordinate approach { (note, however, that many equations in the NP formalism are redundant, see e.g. \cite{Edgar1992} and references therein)}, all differential equations in this formalism are of the first order. Another advantage is that one can also use gauge transformations of the frame in order to simplify the field equations. This is why the formalism is especially powerful when studying algebraically special solutions according to the Petrov classification, since in this case some frame components of the Weyl tensor can be set to zero by choosing an appropriate frame.  

In recent years,   solutions to the higher-dimensional Einstein field equations have attracted a lot of interest. Lot of effort went into   generalizing basic concepts, properties and results of the four-dimensional general relativity to higher dimensions and
there is  growing awareness that higher-dimensional gravity contains qualitatively new physics 
(see e.g. \cite{Emparan:2008eg} and references therein). 
  
Generalization of the Petrov classification and { of} the NP formalism to higher dimensions 
was developed in \cite{Coleyetal04,Milsonetal05}, \cite{Pravdaetal04,OrtPraPra07}, respectively. 
Using these methods, it can be shown that in contrast to four dimensions, the Goldberg-Sachs theorem is not valid in higher dimensions since a multiple Weyl aligned null direction (WAND) in higher-dimensional vacuum algebraically special spacetimes can be shearing \cite{Pravdaetal04,OrtPraPra07}. For example, while in four dimensions expanding vacuum type N and III spacetimes are never shearing, in higher dimensions they are always shearing \cite{Pravdaetal04}. This presence of shear in higher dimensions can substantially complicate the process of solving the field equations. 

In the present paper, we apply the higher-dimensional NP formalism to the study of vacuum spacetimes admitting a non-twisting and (possibly) shearing geodetic multiple WAND and thus belonging to Weyl types II, D, III or N \cite{Coleyetal04,Milsonetal05}.
After introductory remarks and necessary definitions, in section \ref{SecIIDIIIN} we study   dependence of  the metric of the above-mentioned classes of spacetimes on the affine parameter $r$  along null geodesics generated by the multiple WAND.  It is also pointed out that in fact main results of this section also apply  to a special subclass I(a) of the type I. In appropriate coordinates, the $r$-dependence of all components of the metric except of the component $g_{00}$ turns out to be at most quadratic in $r$. The component $g_{00}$ is again quadratic in $r$  for types III and N and more complicated for types II and D. These two cases are thus studied separately.

In section \ref{SecIIIN}  the $r$-dependence of  $g_{00}$ and of the Weyl tensor for types III and N is determined.  It is also shown that when expansion $\theta \not=0$ these spacetimes are singular. In type N  the second order curvature invariant 
$ I=C^{a b c d  ; r s} C_{a m c  n ; r s} C^{t m u n ;v  w} C_{t b u d  ; v w} $ diverges in arbitrary dimension at a point which can be set to $r=0$. Similarly, a first order curvature invariant is  used for type III expanding spacetimes.

In section \ref{SecDII} we determine the $r$-dependence of  $g_{00}$ and of the Weyl tensor for types II and D. Since the problem of solving corresponding differential equations in arbitrary dimension seems to be too complex, we focus on a special case with all non-vanishing eigenvalues of $S_{ij}$ being equal and `antisymmetric' part of the Weyl tensor $\Phi_{ij}^A$ being zero. These assumptions are  satisfied, for example, for all non-twisting Kerr-Schild spacetimes \cite{KSprep}, in particular for Schwarzschild-Tangherlini black holes or corresponding black strings/branes. It also seems to be reasonable to expect that 
the Weyl tensor in the case with distinct eigenvalues of $S_{ij}$ and $\WDA{ij} =0$  will have the same behaviour in the leading order asymptotically thanks to (\ref{res_s}).  

It turns out that the $r$-dependence of $g_{00}$ for Weyl types II and D is determined by an integer $m$ corresponding to the rank of the expansion matrix $S_{ij}$.  In the expanding case,  apart from a quadratic polynomial  in $r$,  $g_{00}$ also contains  a term proportional to $r^{1-m}$ for $m\not=1$ and $\ln r$ for $m=1$. {\footnote{Note that since we do not employ all field equations of the NP formalism, it may in fact turn out that solutions corresponding to the case $m=1$ do not exist.
In four dimensions the case $m=1$ is forbidden by the Goldberg-Sachs theorem.}} Using similar arguments as in \cite{KSprep} it can be shown that in the expanding case the Kretschmann curvature invariant $R_{abcd} R^{abcd}$ diverges for $r=0$ and that it is regular there in the non-expanding case. We also briefly discuss the shear-free case which occurs for $m=0$ (Kundt spacetimes) and for $m=n-2$ (Robinson-Trautman spacetimes). In contrast to the four-dimensional general relativity, in the $m=n-2>2$ case, boost weight $-1$ and $-2$ components of the Weyl tensor necessarily vanish and the spacetime
is thus  of type D in agreement with \cite{PodOrt06}.

In   section \ref{Sec5D}, in order to provide an illustrative example of the use of the higher-dimensional NP formalism,  we focus on solving the full set of the field equations for type N.  To considerably simplify resulting  equations, we make several additional assumptions on the metric and  arrive at an exact  vacuum solution. However, after a coordinate transformation it can be  found that the resulting solution could be  obtained as a direct
product of a  four-dimensional type N Robinson-Trautman metric with an extra dimension. 

The higher-dimensional vacuum Ricci \cite{OrtPraPra07} and Bianchi \cite{Pravdaetal04} equations, extensively used throughout this paper, are given 
in a parallelly propagated frame with a multiple WAND in \ref{sec_ricci} and \ref{appbianchi}, respectively.

\section{Preliminaries}

\subsection{Algebraic classification of the Weyl tensor and Newman-Penrose formalism in higher dimensions}
\label{SecPrelim1}

For convenience, let us  briefly summarize basic aspects of algebraic classification of the Weyl tensor  and the Newman-Penrose formalism in higher dimensions needed in the following sections. More information can be found in original references \cite{Coleyetal04,Milsonetal05} (classification) and \cite{Pravdaetal04,OrtPraPra07} (NP-formalism).
Algebraic classification of the Weyl tensor in higher dimensions was also reviewed in \cite{Coleyrev}.

We introduce a null  frame with  two null vectors $\bm^{(1)}=\bm_{(0)}  =\bl$, $\bm^{(0)}=\bm_{(1)}  ={\bn}$,
and $n-2$ orthonormal spacelike vectors $\bm^{(i)}=\bm_{(i)}$
subject to 
\BE
\fl
\ell^a \ell_a= n^a n_a =\ell^a m^{(i)}_{a}=n^a m^{(i)}_a= 0, \qquad  \ell^a n_a = 1, \qquad
 		m^{(i)a}m^{(j)}_a=\delta_{ij}.\label{normalization}
\EE
The metric reads
\BE
g_{a b} = 2\ell_{(a}n_{b)} + \delta_{ij} m^{(i)}_a m^{(j)}_b. \label{metric}
\EE
Indices $a,b,\dots$ take values from 0 to $n-1$, while $i,j,\dots$ from 2 to $n-1$. Note also that since indices $i,j,\dots$ are raised/lowered by $\delta_{ij}$ there is no need to distinguish between subscripts and superscripts of this type. 

Lorentz transformations are generated by null rotations
\BE
 \hbl=\bl, \qquad \hbn =\bn+z_i\bm{i} -\pul z^2\bl , \qquad \hbm{i} =\vbm{i} -z_i\bl ,
    \label{tr-nullrot}
\EE
with $z^2\equiv z_iz^i$, spins
\BE
 \hbl =  \bl, \qquad \hbn = \bn, \qquad \hbm{i} =  X^{i}_{\ j} \vbm{j} , 
 \label{tr-spins}
\EE
with $X^{i}_{\ j}$ being orthogonal matrices and boosts
\BE
 \hbl = \lambda \bl, \qquad  \hbn = \lambda^{-1} \bn, \qquad \hbm{i} = \vbm{i} .
 \label{tr-boosts}
\EE

If a quantity $q$ transforms under a boost (\ref{tr-boosts}) as $\hat q=\lambda^b q$ 
we say that  $q$ has a boost weight $b$.

The Ricci rotation coefficients $L_{ab}$, $N_{ab}$ and $\Mi_{ab}$ are defined by \cite{Pravdaetal04}
\BE
\fl 
\ell_{a;b}=L_{cd} m^{(c)}_a m^{(d)}_{b} \ , \qquad n_{a;b}=N_{cd} m^{(c)}_a m^{(d)}_{b} \ , \qquad
m^{(i)}_{a;b}=\Mi_{cd} m^{(c)}_a  m^{(d)}_{b}  \ \label{derlnm}
\EE
and their transformation properties  under (\ref{tr-nullrot})-(\ref{tr-boosts}) are given in \cite{OrtPraPra07}. 
These quantities satisfy constraints  
\BEA 
\fl
 L_{0a}=N_{1a}&=&0,\\
 \fl
  N_{0a}+L_{1a}&=&0,\ \ \Mi_{0a} + L_{ia}=0,\ \   \Mi_{1a}+N_{ia}=0,\ \ \Mi_{ja}+\Mj_{ia}=0 . 
\EEA
In four dimensions, $L_{ab}$, $N_{ab}$ and $\Mi_{ab}$ are equivalent to standard complex NP spin coefficients $\kappa$, $\sigma$, $\rho$, etc. (see \cite{OrtPraPra07} for the correspondence).

%Since here we assume $\bl$  to be   geodetic and affinely parameterized, $L_{i0}=0=L_{10}$.
%In following sections we choose a frame that is parallelly propagated and thus also $N_{i0}=0=\Mi_{j0}$. 
%For geodetic $\bl$, $L_{ij}$ can be decomposed  \cite{Pravdaetal04} (cf also \cite{OrtPraPra07})
%nto shear  $\sigma_{ij}$ (tracefree  symmetric part),
%expansion $\theta$ (trace) and twist $A_{ij}$ (antisymmetric part) as
%\BE
%L_{ij}=\sigma_{ij}+\theta\delta_{ij}+A_{ij}.
%\EE
%We will also often denote symmetric part of $L_{ij}$ as expansion matrix $S_{ij}$. Obviously $S_{ij}=\sigma_{ij}+\theta\delta_{ij}$.  

Covariant derivatives along the frame vectors are defined by
\BE
D \equiv \ell^a \nabla_a, \qquad \bigtriangleup  \equiv n^a \nabla_a, \qquad \delta_i \equiv m^{(i)a} \nabla_a .
\label{operators}
\EE

By introducing notation
\BE
T_{\{pqrs\}} =\pul (T_{[ab][cd]} + T_{[cd][ab]}),
\EE
we can decompose the Weyl tensor  and sort its components by boost weight \cite{Milsonetal05}
\BEAH
  C_{abcd} = 
  \overbrace{
    4 C_{0i0j}\, n^{}_{\{a} m^{(i)}_{\, b}  n^{}_{c}  m^{(j)}_{\, d\: \}}}^2  \\
  \nonumber
  +\overbrace{
    8C_{010i}\, n^{}_{\{a} \ell^{}_b n^{}_c m^{(i)}_{\, d\: \}} +
    4C_{0ijk}\, n^{}_{\{a} m^{(i)}_{\, b} m^{(j)}_{\, c} m^{(k)}_{\, d\: \}}}^1  
  \nonumber \\
    \begin{array}{l}
      +4 C_{0101}\, \, n^{}_{\{a} \ell^{}_{ b} n^{}_{ c} \ell^{}_{\, d\: \}} 
\;  + \;  4 C_{01ij}\, \, n^{}_{\{a} \ell^{}_{ b} m^{(i)}_{\, c} m^{(j)}_{\, d\: \}}  \\[2mm]
      +8 C_{0i1j}\, \, n^{}_{\{a} m^{(i)}_{\, b} \ell^{}_{c} m^{(j)}_{\, d\: \}}
   +  C_{ijkl}\, \, m^{(i)}_{\{a} m^{(j)}_{\, b} m^{(k)}_{\, c} m^{(l)}_{\, d\: \}}
    \end{array}
\Biggr\}^0
  \label{eq:rscalars}\\ 
   + \overbrace{
    8 C_{101i}\, \ell^{}_{\{a} n^{}_b \ell^{}_c m^{(i)}_{\, d\: \}} +
    4 C_{1ijk}\, \ell^{}_{\{a} m^{(i)}_{\, b} m^{(j)}_{\, c} m^{(k)}_{\, d\: \}}}^{-1} 
  \nonumber \\
  + \overbrace{
      4 C_{1i1j}\, \ell^{}_{\{a} m^{(i)}_{\, b}  \ell^{}_{c}  m^{(j)}_{\, d\: \}}}^{-2},
\EEAH
where boost weight of various components is indicated by integers (-2,\dots,\ 2). Note that frame components of the Weyl tensor are subject to constraints \cite{Pravdaetal04} following from symmetries of the Weyl tensor
\BEA 
  &&C_{0[i|0|j]} = 0,\nn\\
  \nonumber
  &&C_{0i(jk)}=C_{0ijk}+C_{0kij}+C_{0jki} = 0,\\
  &&C_{ijkl} = C_{\{ijkl\}},\quad C_{ijkl}+C_{iljk}+C_{iklj}=0,\quad 
  C_{01ij}= 2 C_{0[i|1|j]}, \nn\\
  \nonumber
  &&C_{1i(jk)}=C_{1ijk} +C_{1kij} +C_{1jki} = 0, \\
  &&C_{1[i|1|j]} = 0 \label{eq:rcomps}
\EEA
and from its tracelessness
\BEA
  \nn
  && C_{0i0i} =  C_{1i1i} =0,\\
  \nonumber
  &&C_{010i} = C_{0jij},\quad C_{101i} = C_{1jij},\\
     && 2C_{0i1j}=C_{01ij}- C_{ikjk},\quad 
  C_{0101} = -\pul C_{ijij}.\label{eq:rcomps_trace}
\EEA 
We obtain following numbers of independent Weyl tensor frame components of various boost weights  \cite{Pravdaetal04}
\BDM
\fl
\overbrace{2 \left( \frac{n(n-3)}{2} \right) }^{2,-2} + \overbrace{2 \left( \frac{(n-1)(n-2)(n-3)}{3} \right) }^{1,-1} 
+\overbrace{\frac{(n-2)^2 (n-1)(n-3)}{12} + \frac{(n-2)(n-3)}{2}}^0 ,
\EDM
which is in agreement with the number of independent components of the Weyl tensor being
$(n+2)(n+1)n(n-3)/{12}$.

We define  boost order of a tensor {\mbox{\boldmath{$T$}}} to be  boost weight of its leading term. It turns out that boost order of a tensor  depends only on vector $\bl$, being independent on the choice of $\bn$ and $\bm^{(i)}$ \cite{Milsonetal05}. Therefore, given a tensor {\mbox{\boldmath{$T$}}},  preferred null directions may exist for which boost order of  {\mbox{\boldmath{$T$}}} is less then for a generic choice of  $\bl$. Algebraic classification of tensors in higher dimensions \cite{Milsonetal05} is based on existence (and multiplicity) of these preferred null directions in a given spacetime. In case of the Weyl tensor,
 we call them Weyl aligned null directions (WANDs) and spacetime is said to be of principal type G (general) if there are no WANDs, and of principal type I, II, III and N if there are WANDS of multiplicity  1, 2, 3, 4, respectively. Therefore in type I, II, III and N spacetimes all Weyl tensor components with  boost weight higher or equal to 2, 1, 0, -1, respectively can be transformed away by an appropriate choice of the frame vector $\bl$. 
 In some cases one can also set trailing  frame components to zero, and this is the basis of the secondary classification. For instance in type D (principal type II, secondary type {\it ii}), only boost weight zero components are non-vanishing in an appropriately choosen frame.
 In four dimensions principal and secondary classification reduce to the well known Petrov classification.

In agreement with \cite{Pravdaetal04} we introduce notation appropriate for type III and N spacetimes
\BE
\Psi_i\equiv C_{101i},\ \ \ \Psi_{ijk}\equiv \pul C_{1kij},\ \ \ \Psi_{ij}\equiv \pul C_{1i1j},\label{Weyl_III}
\EE
where from (\ref{eq:rcomps}), (\ref{eq:rcomps_trace}) $\Psi_i$, $\Psi_{ijk}$ and $\Psi_{ij}$ satisfy
\BE
\fl
\Psi_i=2\Psi_{ijj},\ \ \ \Psi_{ijk}=-\Psi_{jik},\ \ \ \Psi_{ ijk }+\Psi_{ kij }+\Psi_{ jki }=0,\ \ \ \Psi_{ij}=\Psi_{ji},
\ \ \ \Psi_{ii}=0. \label{cykl}
\EE
Thus, e.g., in type N spacetimes, the Weyl tensor is given by 
\BE
C_{abcd}=8 \Psi_{ij}\, \ell^{}_{\{a} m^{(i)}_{\, b}  \ell^{}_{c}  m^{(j)}_{\, d\: \}}
\EE
and is determined by $\frac{n (n-3)}{2}$ components of the symmetric traceless $(n-2) \times (n-2)$ matrix  $\Psi_{ij}$.

For describing boost weight zero components   of the Weyl tensor 
we will introduce real matrix $\WD{ij} $ as in \cite{PraPraOrt07}
\BE
 \WD{ij} \equiv C_{0i1j} .\label{Psi}
\EE
Then  from (\ref{eq:rcomps}), (\ref{eq:rcomps_trace})
\BE \fl
C_{01ij}=2 C_{0[i|1|j]}=2\WDA{ij} ,\ \  C_{0(i|1|j)}=\WDS{ij} = - \textstyle{\frac{1}{2}} C_{ikjk}, \ \ 
C_{0101}= - \textstyle{\frac{1}{2}} C_{ijij}=\WD{ } ,\label{Psi_SA}
\label{Wcomptsid}
\EE
with $\WDS{ij} $, $\WDA{ij} $, and $\WD{ } \equiv \WD{ii} $ being the symmetric and antisymmetric parts
of $\WD{ij} $ and its trace, respectively. Boost weight zero components   of the Weyl tensor
are thus determined by $\WD{ij} $ and  $C_{ijkl}$.

%-------------------------------------------

\subsection{Spacetimes admitting non-twisting WANDs}
\label{SecPrelim2}

We consider an $n$-dimensional vacuum spacetime admitting  a  non-twisting  geodetic\footnote{Note that in vacuum type N and III  spacetimes a multiple WAND is         always  geodetic \cite{Pravdaetal04}, while in type II and D it is geodetic in the   `generic' case  (see \cite{PraPraOrt07} for details).}
 null congruence generated by  a multiple WAND $\bl$.
Thus $\bl$ is normal and tangent  to 
null hypersurfaces $u=$const 
($g^{ab}u,_a u,_b =0$, $a,b=0\dots\ n-1$) and the WAND
$\ell^a = g^{ab}u,_b$  is thus indeed  geodetic and affinely parameterized,  $\ell^a{;_b}\ell^b=0$.

Similarly as in \cite{NP,PodOrt06},
we choose a coordinate $x^0\equiv u$, a coordinate $x^1\equiv r$, where $r$ is an affine parameter along null geodesics generated  by $\bl$,
and `transverse'
coordinates $x^\alpha$ ($\alpha=2\dots n-1$) labeling  the null
geodesics on hypersurfaces $u=$const and being constant along each geodesic. For the contravariant components of the metric tensor it  follows that 
$g^{01}=1$, $g^{00}=0=g^{0\alpha}$.
Then the frame $\bl$, ${\bn}$,
and  $\bm^{(i)}=\bm_{(i)}$ satisfying (\ref{normalization}) can be given as
\BEA
\ell^a & =[0,1,0,\dots,0], \ \ \  & \ell_a \ \  =[1,0,\dots,0],\label{l}\\
n^a & =[1,U,X^\alpha], \ \ \   \quad\quad  & n_a \ \  =[V,1,Y_\alpha],\label{n}\\
m^a_{(i)} & =[0,\omega_{i},\xi^\alpha_{i}],\ \ \  \quad & m_a^{(i)} =[\Omega^{i},0,\eta^{i}_\alpha].\label{m}
\EEA
Equations (\ref{normalization}) 
 imply
\BEA
0\ \ =U+V+X^\alpha Y_\alpha,\label{norm_nn}\\
0\ \ =\omega_{i}+\xi^\alpha_{i}Y_\alpha,\label{norm_nm}\\
0\ \ =\Omega^{i}+\eta_\alpha^{i}X^\alpha,\label{norm_nm2}\\
\delta_{i}^j=\xi^\alpha_{i}\eta^{j}_\alpha.\label{norm_mm}
\EEA
 By multiplying (\ref{norm_mm}) by $\eta^i_\beta$ we get 
$\delta_{i}^j\eta^i_\beta=\eta^j_\beta=(\eta^i_\beta\xi^\alpha_{i})\eta^{j}_\alpha$
which gives
\BE
\delta_{\beta}^\alpha=\xi^\alpha_{i}\eta^{i}_\beta .\label{norm_delta}
\EE

%The Ricci rotation coefficients $L_{ab}$, $N_{ab}$ and $\Mi_{ab}$ are defined by \cite{Pravdaetal04}
%\BE
%\fl 
%\ell_{a;b}=L_{cd} m^{(c)}_a m^{(d)}_{b} \ , \qquad n_{a;b}=N_{cd} m^{(c)}_a m^{(d)}_{b} \ , \qquad
%m^{(i)}_{a;b}=\Mi_{cd} m^{(c)}_a  m^{(d)}_{b}  \ \label{derlnm}
%EE
%and satisfy  
%\BEA 
%\fl
% L_{0a}=N_{1a}&=&0,\\
% \fl
%  N_{0a}+L_{1a}&=&0,\ \ \Mi_{0a} + L_{ia}=0,\ \   \Mi_{1a}+N_{ia}=0,\ \ \Mi_{ja}+\Mj_{ia}=0 . 
%\EEA
Since $\bl$ is  geodetic and affinely parameterized, $L_{i0}=0=L_{10}$.
Let us choose a frame that is parallelly propagated, i.e. $N_{i0}=0=\Mi_{j0}$. 
For geodetic $\bl$, $L_{ij}$ can be decomposed  \cite{Pravdaetal04} (cf also \cite{OrtPraPra07})
into shear  $\sigma_{ij}$ (trace-free  symmetric part),
expansion $\theta$ (trace) and twist $A_{ij}$ (antisymmetric part) as
\BE
L_{ij}=\sigma_{ij}+\theta\delta_{ij}+A_{ij}.
\EE
We will also often denote symmetric part of $L_{ij}$ as expansion matrix $S_{ij}$. Obviously $S_{ij}=\sigma_{ij}+\theta\delta_{ij}$.

%Covariant derivatives along the frame vectors are defined by
%\BE
%D \equiv \ell^a \nabla_a, \qquad \bigtriangleup  \equiv n^a \nabla_a, \qquad \delta_i \equiv m^{(i)a} \nabla_a .
%\EE
When acting on a function $f$, the operators (\ref{operators}) 
and their commutators \cite{Coleyetal04vsi} can be expressed as
\BE
 D 
=\partial_r,\ \ \ 
\T=\partial_u+U\partial_r+X^\alpha\partial_\alpha,\ \ \ 
\delta_i=\omega_{i}\partial_r+\xi^\alpha_{i}\partial_\alpha
\EE
and
\BEA
\fl
(\T D - D \T)f &= L_{11} Df  
+ L_{i1} \delta_i f
, \label{commutators-1}\\
\fl
(\delta_i D - D \delta_i)f &= L_{1i} Df+ L_{ji}\delta_j f, \label{commutators-2}\\
\fl
(\delta_i \T - \T \delta_i)f & = N_{i1} Df + (L_{i1}-L_{1i}) \T f+ (N_{ji}-\Mi_{j1}) \delta_j f, \label{commutators-3}\\
\fl
(\delta_i \delta_j - \delta_j \delta_i)f &= (N_{ij}-N_{ji}) D f+ (L_{ij}-L_{ji})\T f
 + (\Mj_{ki}-\Mi_{kj})\delta_k f.\label{commutators-4}
\EEA

Apart from Bianchi equations \cite{Pravdaetal04} and Ricci equations \cite{OrtPraPra07} we need relations between metric
components and the Ricci rotation coefficients. Such relations may be obtained by applying the commutators 
(\ref{commutators-1})--(\ref{commutators-4}) on coordinates $u$, $r$, $x^\alpha$.
For $f=u$,  
(\ref{commutators-3}) and (\ref{commutators-4}) imply
\BEA
0&=&L_{i1}-L_{1i}\label{com_u_L1i},\\
0&=&L_{ij}-L_{ji}\label{com_u_Lij}.
\EEA
For $f=r$, (\ref{commutators-1})--(\ref{commutators-4}) lead to
\BEA
\quad\quad -DU&=L_{11}+L_{i1}\omega_{i}\label{com_r_U},\\
\quad\quad -D\omega_{i}&=L_{1i}+L_{ji}\omega_{j},\label{com_r_om}\\
\delta_i U-\T\omega_{i}&=N_{i1}
+(N_{ji}-\Mi_{j1})\omega_{j},\label{com_r_delU}\\
\delta_i\omega_{j}-\delta_j\omega_{i}&=N_{ij}-N_{ji}
+(\Mj_{ki}-\Mi_{kj})\omega_{k},\label{com_r_delom}
\EEA
and for $f=x^\alpha$, (\ref{commutators-1})--(\ref{commutators-4}) give
\BEA
\quad\quad
-DX^\alpha&=L_{j1}\xi^\alpha_{j},\label{com_x_X}\\
\quad\quad
-D\xi^\alpha_{i}&=L_{ji}\xi^\alpha_{j},\label{com_x_xi}\\
\delta_i X^\alpha -\T\xi^\alpha_{i}&=
(N_{ji}-\Mi_{j1})\xi^\alpha_{j},\label{com_x_delX}\\
\delta_i\xi^\alpha_{j}-\delta_j\xi^\alpha_{i}&=
(\Mj_{ki}-\Mi_{kj})\xi^\alpha_{k}.\label{com_x_delxi}
\EEA

%-----------------------------------------------------

\subsection{Indices}
\label{indices}

For convenience let us   summarize types of the indices used  throughout this paper.
Apart from indices
$a,b,\dots =0,\dots , n-1$, and $i,j,\dots =2,\dots ,n-1$ introduced in section \ref{SecPrelim1}, we also 
introduce indices $\alpha, \beta = 2,\dots ,n-1$ numbering spacelike coordinates and corresponding components
in section \ref{SecPrelim2}. 

In four dimensions, the expansion matrix $S_{ij}$ is  of rank 2 in the expanding case due to the Goldberg-Sachs theorem. 
However, in higher dimensions
 $m \leq n-2$, where $m$ is rank of  $S_{ij}$.  In next sections  we will often need to distinguish between indices corresponding  to non-vanishing ($o,p,q,s =2,\dots,m+1$) and  vanishing  ($v,w,y,z  = m+2,\dots ,n-1$) eigenvalues of $S_{ij}$.

In following calculations it also turns out to be practical to  modify  Einstein's summation convention for indices $o,p,q,s$:
in an expression there is  summation over repeated indices if  there are two indices without brackets among them 
(thus e.g. in $\eta_\alpha^{p0}\eta_\beta^{p0}X^{\beta 0}(r+a_{(p)})^2$ 
there is summation
over $p$ 
while in $\WD{pq} s_{(p)}$ we do not sum over $p$).

%-----------------------------------------------------------

\section{Radial integration  for non-twisting vacuum Weyl type II, D, III, N spacetimes}
\label{SecIIDIIIN}

In the present paper we study  $r$-dependence of the metric functions, the Ricci rotation coefficients and the Weyl tensor, which,  however, is in general
different  for various algebraic types. 
In order to avoid repetition, in this section we focus on those metric functions  and Ricci rotation coefficients 
that have the same $r$-dependence for all  algebraic types studied.
Note that in contrast to section \ref{SecDII}, here we do not assume that all non-vanishing eigenvalues of the expansion matrix $S_{ij}$ are equal.

Without loss of generality we choose the frame (\ref{l})--(\ref{m}) in such a way that  
$S_{ij}$  is diagonal, $S_{ij}=$diag$\{ s_{(2)},\dots,\ s_{(m+1)},0, \dots, 0\}$, where $m$ denotes number of non-zero eigenvalues of $S_{ij}$.  
As it is shown in \cite{prep2},  
%it follows from equation (11g,\cite{OrtPraPra07}) that    DOPLNIT@@@@@@@@@@@@@@@@@@@@@@@@@@@@@
this assumption is compatible with the frame being parallelly transported.
As mentioned in section \ref{indices}, 
%indices $i,j,k \dots$ run from $2$ to $n-1$. We will often need to distinguish between values $2\dots m+1$ corresponding to %non-vanishing eigenvalues of $S_{ij}$ and values $m+2\dots n-1$ corresponding to vanishing eigenvalues of $S_{ij}$.  
 %we introduce %the following notation:
indices $o$, $p$, $q$, $s$ corresponding to non-vanishing eigenvalues of $S_{ij}$ run from $2$ to $m+1$ and 
indices $v$, $w$, $y$, $z$ corresponding to vanishing eigenvalues of $S_{ij}$ run from $m+2 \dots n-1$.

In our case,  from Ricci equations (\ref{11g}) %,\cite{OrtPraPra07})  
%reduce to $DL_{ij}=-L_{ik}L_{kj}$. 
for non-vanishing
eigenvalues of $S_{ij}$, $s_{(p)}\not= 0$, it follows 
\BE
s_{(p)}=\frac{1}{r+a_{(p)}^0},\label{res_s}
\EE
where $a_{(p)}^0$ is an arbitrary function of $u$ and $x^\alpha$, independent on $r$. 
Similarly, throughout this paper, the superscript `$0$' will   suggest that the function under consideration does not depend on $r$.

Ricci equations (\ref{11b})=(\ref{11e}), $DL_{1i}=-L_{1i}s_{(i)}$, 
lead to 
\BE
L_{1p}=\frac{l_{1p}^0}{r+a^0_{(p)}},\ \ \ 
L_{1w}=l_{1w}^0.\label{L1is0}
\EE

There is still freedom to perform a null rotation with fixed $\bl$ %\cite{Milsonetal05,OrtPraPra07} 
(\ref{tr-nullrot}).
%\BE
%\fl
% \hbl=\bl, \qquad \hbn =\bn+z_i\bm^{(i)} -\pul z^2\bl , \qquad \hbm{i} =\bm^{(i)} -z_i\bl,\ \ z^2=z_iz_i .
%    \label{nullrot}
%\EE
To preserve parallel propagation of the frame, $z_i$ is subject to 
\BE
Dz_i=0.\label{Dz}
\EE 
Choosing $z_p=-l_{1p}^0$,  we can set $L_{1p}$  to zero by (see \cite{OrtPraPra07})
\BE
\hat L_{1p}=L_{1p}+z_jL_{ji}=0.
\EE
In what follows we omit the hat symbol. 
Note that parameters $z_w$ can be used to further simplify the metric, e.g. one can set $\omega_w^0$ to zero
as in section \ref{sec_spec-vanishL1i} and section \ref{Sec5D}.

From  Ricci equations (\ref{11n}), reduced to $D\Mj_{ki}=-\Mj_{ki}s_{(i)}$, 
(\ref{com_x_xi}) and  (\ref{com_r_om}),  we obtain 
\BEA
\Mj_{kp}&=\frac{\mjo{kp} }{r+a_{(p)}^0},\ \ \ \Mj_{kw}&={    \mjo{kw} } ,
\label{Mijks}\\
\xi^\alpha_{p}&=\frac{\xi^{\alpha 0}_{p}}{r+a_{(p)}^0},\ \ \ \quad\xi^\alpha_{w}&=\xi^{\alpha 0}_{w},
\label{xiijs}\\
\omega_ {p}&=
\frac{\omega_ {p}^0}{r+a_{(p)}^0}, \ \ \quad\omega_ {w}&=-l_{1w}^0r+\omega_ {w}^0,
\label{omis}
\EEA
respectively
and
from (\ref{com_x_X})
\BE
X^\alpha=  -l^0_{1w}\xi^{\alpha 0}_ {w}r+ X^{\alpha 0}.\label{X}
\EE

To compute the covariant components of the metric one has to solve (\ref{norm_nn})--(\ref{norm_mm})
for $\eta^i_\alpha$, $Y_\alpha$, $\Omega^p$, $V$.
From (\ref{norm_nn})--(\ref{norm_mm})
also using   (\ref{norm_delta}) and (\ref{xiijs})--(\ref{X}), it follows
\BEA
\fl
\eta^p_\alpha&=&\eta^{p0}_\alpha (r+a_{(p)}),\ \ \ \eta^w_\alpha=\eta^{w0}_\alpha ,\label{eta}\\
\fl
Y_\alpha &=&-\eta^i_\alpha \omega_i=
l_{1w}^0\eta^{w0}_\alpha r -(\eta^{p0}_\alpha \omega^0_p+\eta^{w0}_\alpha \omega_w^0),\label{YN}\\
\fl
\Omega^{p}&=&-\eta_\alpha^{p}X^\alpha=
-\eta_\alpha^{p0}X^{\alpha 0}(r+a_{(p)}),\\
\fl
\Omega^{w}&=&-\eta_\alpha^{w}X^\alpha=
l_{1w}^0r-\eta_\alpha^{w0}X^{\alpha 0},
\label{Omega_w}\\
\fl
V&=&-U+l_{1w}^0l_{1w}^0 r^2-(\omega_{w}^0+\eta^{w0}_\alpha X^{\alpha 0})l_{1w}^0 r
+X^{\alpha 0}(\eta^{p0}_\alpha \omega^0_p+\eta^{w0}_\alpha \omega_w^0).
\label{V}
\EEA
As will be discussed below, the $r$-dependence of the function $U$ has to be studied separately for types II, D and III, N.

The covariant components of the metric tensor (cf (\ref{metric})) thus read 
\BEA
\fl
g_{11}&=&0,\ \ \  g_{01}=1,\ \ \ g_{1\alpha}=0,\ \ \ \label{metric11}\\
\fl
g_{00}&=&2V+{\Omega^i}\Omega^i =2V
+\eta_\alpha^{p0}X^{\alpha 0}\eta_\beta^{p0}X^{\beta 0}(r+a_{(p)})^2
\nonumber\\ \fl &&
+(l_{1w}^0r-\eta_\alpha^{w0}X^{\alpha 0})(l_{1w}^0r-\eta_\beta^{w0}X^{\beta 0}), \label{metric00}\\ 
\fl
g_{0\alpha}&=&Y_\alpha+{{\Omega^j}}{\eta^{j}_{\alpha}}\nonumber\\
\fl
&=&-\eta_\alpha^{p0}\eta_\beta^{p0}X^{\beta 0}(r+a_{(p)})^2+2l_{1w}^0\eta^{w0}_\alpha r
-(\eta^{p0}_\alpha\omega^0_p+\eta^{w0}_\alpha\omega_w^0+\eta_\alpha^{w0}\eta_\beta^{w0}X^{\beta 0})\nonumber\\
\fl
&=&\gamma_{\alpha}^{2}r^2+\gamma_{\alpha}^{1}r+\gamma_{\alpha}^{0},\ \ \  \label{IIg0i}\\
\fl
g_{\alpha\beta}&=&\eta_\alpha^k\eta_\beta^k=\eta_\alpha^{p0}\eta_\beta^{p0}(r+a_{(p)})^2+\eta_\alpha^{w0}\eta_\beta^{w0}
=\gamma_{\alpha\beta}^{2}r^2+\gamma_{\alpha\beta}^{1}r+\gamma_{\alpha\beta}^{0}
,\label{cov.metric_D,III,N}
\EEA
therefore {\it the vacuum  metric  with a non-twisting geodetic multiple WAND has the form}
\BE
\fl
{\mbox{d}}s^2\!\!=g_{00}{\mbox{d}}u^2\!+2{\mbox{d}}u{\mbox{d}}r+2\!\left(\gamma_{\alpha}^{2}r^2+\gamma_{\alpha}^{1}r+\gamma_{\alpha}^{0}\right)\!{\mbox{d}}u{\mbox{d}}x^\alpha+\left(\gamma_{\alpha\beta}^{2}r^2+\gamma_{\alpha\beta}^{1}r+\gamma_{\alpha\beta}^{0}\right)\!{\mbox{d}}x^\alpha{\mbox{d}}x^\beta, 
\label{typeIImetric}
\EE
{\it where functions $\gamma_{\alpha\beta}^{N}$ and $\gamma_{\alpha}^{N}$, $N=0,1,2$, introduced in (\ref{IIg0i}), (\ref{cov.metric_D,III,N}) do not depend on $r$. }

Differentiating  equation (\ref{com_r_U}) with respect to $r$ and using (\ref{com_r_om}), (\ref{V}) and  the Ricci equation  (\ref{11a})  for $L_{11}$,
%\BE
%DL_{11}=-L_{1i}L_{1i}-C_{0101},\label{eqL11}
%\EE
we arrive to 
\BE
C_{0101} = - V_{,rr}.\label{C0101Vrr}
\EE
Consequently, for type III and N spacetimes (where $C_{0101}$ has to vanish) $V$ is  linear in $r$, while for type II and D spacetimes the $r$-dependence of $V$ (and hence of $U$) can be more complicated. 
Types II, D and III, N  will be thus discussed separately in the following sections.
Note that for deriving the metric (\ref{typeIImetric}) only assumptions $C_{0i0j}=C_{010i}=0$ on the Weyl tensor are necessary and it was not necessary to assume $C_{0kij}=0$.
Therefore, the metric (\ref{typeIImetric}) also applies  to the special class of type I  spacetimes with $C_{010i}=0$
 denoted by I(a) in \cite{Coleyetal04}.
As for the Ricci tensor, in fact  up to now we have  assumed only $R_{00}=R_{0i}=0$.

Note that it was shown that for type III and N expanding vacuum spacetimes $m=2$ in arbitrary dimension and that $s_{(2)}=s_{(3)}$ \cite{Pravdaetal04}.
If all non-vanishing eigenvalues of  $S_{ij}$ are equal, i.e. from (\ref{res_s}) $s_{(p)}=1/(r+a^0(u,x^\alpha))$ for all $p$, one 
 can perform a coordinate transformation \cite{PodOrt06} that leaves unchanged  null hypersurfaces
$u=$const and preserves the affine character of the parameter $r$:  
\BE
\tilde r= r+a^0(u,x^\alpha).
\EE
Then from Ricci equations (\ref{11k}) (for $i=k=q$, $j=p$)
\BE
\omega_p^0=0.
\EE
In the following, for simplicity we omit the tilde symbol over $r$ and
over absolute terms, such as $\omega^0_{w}$, $X^{\alpha 0}$, $l_{11}^0$, $U^0$, $\mio{j1} $, $n_{i1}^0$.

%----------------------------------------------------------------

\section{Type III, N }
\label{SecIIIN}

In this section, vacuum type III and N spacetimes are considered and $r$-dependence of the remaining metric component $g_{00}$, the Ricci rotation coefficients $L_{11}$, $N_{ij}$,
and $\Mi_{j1}$ and the Weyl tensor are determined. 
These spacetimes are either non-expanding (Kundt class) with $m=0$ or expanding with $m=2$ \cite{Pravdaetal04}, where, in appropriate coordinates $s_{(2)}=s_{(3)}=1/r$, as mentioned above.

From Ricci equations 
(\ref{11a}) and (\ref{com_r_U})  it follows
\BEA
L_{11}&=&
- {l^0_{1w}}l^0_{1w}r+l_{11}^{0}, \label{L11N}\\
U&=&
l^0_{1w} {l^0_{1w}}r^2-( l_{11}^{0}+ l_{1w}^0 \omega_ {w}^0)r+ U^0. \label{UN}
\EEA

For future reference let us note that
one can still perform 
a null rotation with fixed $\bl$ (\ref{tr-nullrot}) with $z_p=0$ for $p=2,3$, $z_w$ arbitrary
and subject to (\ref{Dz})  
\BEA
\fl
\hat L_{1p}=0 ,\ 
\hat L_{1w}=L_{1w},\\
\fl
\hat \omega_ {p}=0 
,\ 
\hat \omega_ {w}=-l_{1w}^0r+\omega_ {w}^0-z_w
=-l_{1w}^0r+\hat \omega_ {w}^0,\label{rot_om} \\
\fl
\hM{j}{k}{i}=
\Mj_{ki}+2z_{[k}L_{j]i}, \\
\fl
\hat \xi^\alpha_{i}=\xi^{\alpha}_{i}, \\
\fl
\hat L_{11}\!\!=L_{11}\!\!+z_i(L_{1i}+L_{i1})+z_iz_jL_{ij}=- l^0_{1w}l^0_{1w}r+ l_{11}^{0}+2z_w l_{1w}^0
=- l^0_{1w}l^0_{1w}r+ \hat l_{11}^{0}\ \ \ \ \ \\
\fl
\hat U={l^0_{1w}}l^0_{1w}r^2
-\left[ l_{11}^{0}+l_{1w}^0(\omega_ {w}^0+z_w)\right]r
+ U^0+z_w \omega_ {w}^0-\pul
 {z_{w}z_{w}}\nonumber\\
\fl \ \ \ \  = l^0_{1w}l^0_{1w}r^2
-(\hat l_{11}^{0}+l_{1w}^0\hat\omega_ {w}^0)r+\hat U^0. \label{rorl}
\EEA
By choosing appropriate $z_w$, $w=4,\cdots n-1$, one can simplify $\omega_ {w}$, $U$ or $l_{11}$
(see sections \ref{sec_spec-vanishL1i} and \ref{Sec5D}).

From Ricci equations (\ref{11j}), (\ref{11m})  
\BEA
N_{ip}= \frac{n_{ip}^0}{r},\ \ \  
N_{iw}=n_{iw}^0,\label{NN}\\
\Mi_{j1}=-\mio{jw} l_{1w}^0 r+\mio{j1} .\label{PIII_Mij1}
\EEA

Let us conclude this section by writing down the metric for the Weyl types III, N. 
From (\ref{V}), using (\ref{UN}), we arrive at
\BE
V=( l_{11}^{0}-l_{1v}^0\eta^{v0}_\alpha X^{\alpha 0})r- U^0+X^{\alpha0}
\eta^{w0}_\alpha\omega_w^0.
\EE
Substituting the metric component
\BEA
\fl
g_{00}&=&2V+{\Omega^i}\Omega^i 
=(\eta_\alpha^{p0}\eta_\beta^{p0}X^{\alpha 0}X^{\beta 0}+l_{1w}^0l_{1w}^0)r^2+2r[ l_{11}^{0}-2l_{1v}^0\eta^{v0}_\alpha X^{\alpha 0}]
- 2U^0
\nonumber\\ \fl
&&+2X^{\alpha 0} 
\eta^{w0}_\alpha\omega_w^0+\eta_\alpha^{w0}\eta_\beta^{w0}X^{\alpha 0}X^{\beta 0}
=\gamma^{2}r^2+\gamma^{1}r+\gamma^{0},\label{PIII_g00}
\EEA
into (\ref{metric00}), from (\ref{typeIImetric})
we find that {\it vacuum type III or N metric with non-twisting multiple WAND has the form}
\BE
\fl
{\mbox{d}}s^2\!\!=\!\!(\gamma^{2}r^2+\gamma^{1}r+\gamma^{0}){\mbox{d}}u^2\!
+\!2{\mbox{d}}u{\mbox{d}}r\!
+\!2(\gamma_{\alpha}^{2}r^2+\gamma_{\alpha}^{1}r+\gamma_{\alpha}^{0}){\mbox{d}}u{\mbox{d}}x^\alpha\!
+\!(\gamma_{\alpha\beta}^{2}r^2+\gamma_{\alpha\beta}^{0}){\mbox{d}}x^\alpha{\mbox{d}}x^\beta,\label{metricIII,N}
\EE
{\it where the functions $\gamma^N$, $\gamma_{\alpha}^{N}$ and $\gamma_{\alpha\beta}^{N}$, $N=0,1,2$, are introduced 
in  (\ref{PIII_g00}), (\ref{IIg0i}) and (\ref{cov.metric_D,III,N}), respectively.} 

In fact to derive the metric (\ref{metricIII,N})
only the following assumptions on the Ricci tensor have been made: $R_{00}=R_{0i}=2R_{01}-R/(n-1)=0$.

Note that in the non-expanding case, i.e. for $m=0$, $\gamma_{\alpha}^{2}$ and $\gamma_{\alpha\beta}^{2}$ vanish (see (\ref{IIg0i}), (\ref{cov.metric_D,III,N}))
and  the metric (\ref{metricIII,N}) is compatible with  higher-dimensional Kundt metrics given in \cite{Coleyetal03,Coleyetal06}.
In the expanding case, i.e. $m=2$, the metric (\ref{metricIII,N}) is compatible with four-dimensional vacuum type III and N
Robinson-Trautman solutions (see ,e.g., \cite{Stephanibook}) and with  direct products of these metrics with a flat space.

In the following sections we study $r$-dependence of the Weyl tensor separately for types N and III.

%-------------------------------------------
\subsection{The Weyl tensor for type N}
\label{sec_WeylN}

In this section $r$-dependence of the remaining quantities entering the Ricci and Bianchi equations  
is derived
for vacuum type N spacetimes. 
In an appropriately chosen frame there are only Weyl components
of boost weight $-2$, $\Psi_{ij}\equiv \pul C_{1i1j}$. As was shown in \cite{Pravdaetal04}, $\Psi_{ij}$ can be diagonalized
together with $S_{ij}$ and admits a form $\Psi_{ij}=$diag.$\{ p,\ -p,\ 0,\ \dots 0\}$.
Similarly as in \cite{prep2}, it can be shown 
%NEMAZAT
%from  Ricci equations (11g,\cite{OrtPraPra07}),  reduced to $DL_{ij}=-L_{ik}L_{kj}$,
%and from (B4=23,\cite{Pravdaetal04}), $D\Psi_{ij}=-\Psi_{ik}L_{kj}$, DOPLNIT@@@@@@@@@@@@@@@@@@@@@@
%
%%
%
that the condition  of both $\Psi_{ij}$ and $S_{ij}$ being diagonal 
is compatible with the frame being parallelly propagated.

Equations (\ref{11f})  
  and  (\ref{B4}) lead to
\BEA
N_{i1} = -(n_{iw}^0l_{1w}^0)r+ n_{i1}^0,\\
p=\textstyle{\frac{p^0}{r}}
.\label{pN}
\EEA

As was shown in \cite{Coleyetal04vsi} the curvature invariant
\BE
 I_{N} \equiv C^{a_1 b_1 a_2 b_2  ; c_1 c_2} C_{a_1 d_1 a_2  d_2 ; c_1 c_2} C^{e_1 d_1 e_2 d_2 ;f_1 f_2} C_{e_1 b_1 e_2 b_2  ; f_1 f_2} 
\EE
reduces for non-twisting type N vacuum spacetimes to
\BE
I_{N}=36 (n-2)^8 \theta^8 (\Psi_{ij} \Psi_{ij})^2.
\EE
$I_{N}$ clearly diverges at $r=0$ in the expanding case and therefore a curvature singularity is located there.
The non-expanding (Kundt) case belongs to VSI spacetimes \cite{Coleyetal04vsi}, i.e. spacetimes with vanishing curvature invariants of all orders, and therefore curvature invariants cannot be used for locating possible singularities.

%---------------------------------------------------------------

\subsection{The Weyl tensor for type III} 
\label{sec_III}

Now let us examine $r$-dependence of the Weyl tensor for type III vacuum spacetimes.
In an appropriately chosen frame, there are only Weyl tensor components
of boost weight $-1$ and $-2$, i.e. $\Psi_{i}$, $\Psi_{ijk}$ and $\Psi_{ij}$, respectively
(see (\ref{Weyl_III}),  (\ref{cykl})).

Bianchi equations (\ref{B1}), (\ref{B9}) and (\ref{B4})   read (note that in our case (\ref{B6}) is equivalent to (\ref{B9}))
\BEA
\quad\quad\quad\quad
D\Psi_i&=-2\Psi_e L_{ei}=-2\Psi_i s_{(i)},\label{Bb1}\\
\quad\quad\quad
D\Psi_{jki}&=\Psi_{kei}L_{ej}-\Psi_{jei}L_{ek}=-\Psi_{jki}(s_{(j)}+s_{(k)}),\label{Bb9}\\
2D\Psi_{ij}-\delta_j \Psi_{i}&=2\Psi_{jei}L_{e1}-2\Psi_{ie}L_{ej}+\Psi_{e}\Me_{ij}.\label{Bb4}
\EEA
Equations (\ref{Bb1}), (\ref{Bb9}) imply
\BEA
\quad
\Psi_p&=\frac{\Psi_p^0}{r^2}, 
\quad \quad\Psi_w&=\Psi_w^0,\label{PIII-Psi_p,w}\\
\Psi_{wvi}&=\Psi_{wvi}^0, \ \ 
\Psi_{pwi}&=\frac{\Psi_{pwi}^0}{r},\ \ \ 
\Psi_{pri}=\frac{\Psi_{pri}^0}{r^2}.\label{PIII-Psi_ijk}
\EEA
From (\ref{PIII-Psi_p,w}), (\ref{PIII-Psi_ijk}) and (\ref{cykl}) it follows
\BE
\fl
\Psi_{prw}=\Psi_{wvp}=0,\ \ \
\Psi_{wrp}=\Psi_{wpr},\ \ \
\Psi_{pvw}=\Psi_{pwv}, \ \ \ \Psi_{pww}=0=\Psi_{wpp}.
\EE

Note that some of the Bianchi identities reduce to 
algebraical equations, studied in detail in \cite{Pravdaetal04}.
Here we use results of \cite{Pravdaetal04} to simplify the Weyl tensor
(\ref{PIII-Psi_p,w}), (\ref{PIII-Psi_ijk}). 
Namely, 
equations (54) in \cite{Pravdaetal04} for  ($i=w$, $j=v$, $k=p$) lead to 
\BE
\Psi_{pwv}=0 
\EE
and  for $i,j,k=v,w,z$ in the expanding case $\theta\not=0$ equations (58) in \cite{Pravdaetal04} give
\BE
\Psi_{vwz}=0\ \ \Rightarrow\ \ \Psi_w=0.\label{expand_Psi_www}
\EE

To summarize: non-vanishing boost weight $-1$ Weyl tensor components for $\theta\not=0$ are (cf (C.20) in \cite{Pravdaetal04})
\BEAH
\quad
\Psi_2&=2\Psi_{233}={\textstyle{\frac{\Psi_2^0}{r^2}}},\quad\quad\quad
 \Psi_3&=2\Psi_{322}={\textstyle{\frac{\Psi_3^0}{r^2}}},\\
\Psi_{w22}&=-\Psi_{w33}={\textstyle{\frac{\Psi_{w22}^0}{r}}},\ \ \ 
\Psi_{w23}&=\Psi_{w32}={\textstyle{\frac{\Psi_{w23}^0}{r}}},
\EEAH
while for the non-expanding case $\Psi_w=\Psi_w^0$ and $ \Psi_{wvz}=\Psi_{wvz}^0 $. 

From equations (\ref{Bb4}) 
in the non-expanding case $\theta=0$  the boost weight $-2$
components of the Weyl tensor are 
\BE
\Psi_{wv}={\textstyle{\frac{r}{2}}}\left( \xi_v^{\alpha 0}\Psi_{w,\alpha}^0+2\Psi_{vzw}^0l_{1z}^0+\Psi_z^0\mzo{wv} \right)+\Psi_{wv}^0,
\EE
while in the expanding case with (\ref{expand_Psi_www})
\BEA
\Psi_{wv}=\Psi_{wv}^0-{\textstyle{\frac{1}{2r}}}\Psi_{p}^0 \mpo{wv} ,\\
\Psi_{wp}={\textstyle{\frac{1}{r}}}\Psi_{wp}^0 -{\textstyle{\frac{1}{2r^2}}}\Psi_{q}^0\mqo{wp} \nonumber\\
=\Psi_{pw}=\Psi_{pw}^0-{\textstyle{\frac{1}{2r}}}(2l_{1w}^0\Psi_p^0+\xi_w^{\alpha 0}\Psi_{p,\alpha}^0+\Psi_q^0\mqo{pw} )
+{\textstyle{\frac{1}{2r^2}}}\omega_w^0\Psi_p^0,\\
\Psi_{pq}=-\Psi_{wqp}^0l_{1w}^0+{\textstyle{\frac{\Psi_{pq}^0}{r}}}
-{\textstyle{\frac{1}{2r^2}}}(\Psi_{o}^0\moo{pq} +\xi^{\alpha 0}_q\Psi_{p,\alpha}^0) 
.
\EEA
Considering $\Psi_{ij}=\Psi_{ji}$, we get
\BEA
\quad\quad\quad\quad\quad\quad
\Psi_{wv}^0&=\Psi_{vw}^0,\ \ \ \Psi_{pw}^0=0,\ \ \ \Psi_{pq}^0=\Psi_{qp}^0,\\
\quad\quad\quad\quad\quad\quad
 \Psi_{wp}^0&=-{\textstyle{\frac{1}{2}}}(2l_{1w}^0\Psi_p^0+\xi_w^{\alpha 0}\Psi_{p,\alpha}^0+\Psi_q^0\mqo{pw} ),\\
\quad\quad\quad\quad
\Psi_{p}^0 \mpo{wv} &=\Psi_{p}^0 \mpo{vw} ,\\
\quad\quad\quad\quad
\Psi_{q}^0\mqo{wp} & =-\omega_w^0\Psi_p^0,\\
\Psi_{o}^0\moo{pq} +\xi^{\alpha 0}_q\Psi_{p,\alpha}^0
&=\Psi_{o}^0\moo{qp} +\xi^{\alpha 0}_p\Psi_{q,\alpha}^0.
\EEA

From (\ref{11f}) one can also determine 
the remaining Ricci rotation coefficients 
\BE
\fl
N_{w1}=(-n_{wv}^0l_{1v}^0+\Psi_w^0\delta_{m0})r+n_{w1}^0,\ \ \ 
N_{p1}=-n_{pv}^0l_{1v}^0r+n_{p1}^0-{\textstyle{\frac{\Psi_{p}^0}{r}}}.
\EE

As was shown in \cite{Coleyetal04vsi}, the curvature invariant
\BE
I_{{III}} = C^{a_1 b_1 a_2 b_2;e_1} C_{a_1 c_1 a_2 c_2;e_1} C^{d_1 c_1 d_2 c_2;e_2} C_{d_1 b_1 d_2 b_2;e_2}
\EE
can be expressed as (74) in \cite{Coleyetal04vsi}\footnote{Equation (\ref{Inv3old}) is expressed using the notation of \cite{Coleyetal04vsi}, while in (\ref{Inv3}) it is rewritten in terms of the quantities introduced in the present paper. Note also there is a misprint in equation (74) in \cite{Coleyetal04vsi}. It was obtained in Maple using definition $\psi={\Psi_i \Psi_i}$, while standard definition, used also in \cite{Coleyetal04vsi} and in the present paper,  is $\psi^2={\Psi_i \Psi_i}$. Therefore $\psi$  in equation (74) from \cite{Coleyetal04vsi} has to be replaced by $\psi^2$.}
\BEA
\fl
I_{{III}} &=& 64 S^4 
\left[ 9 \psi^4 +27 \psi^2
(\cO_{PP} + \cO_{PF}) + 28 (\cO_{PP} + \cO_{PF})^2 \right]  \label{Inv3old} \\
\fl
&=&
4(n-2)^4\theta^4\left[ 9 \psi^{4} +27 \psi^{2}
({\Psi_{w22}}^2 + {\Psi_{w23}}^2) + 28({\Psi_{w22}}^2 + {\Psi_{w23}}^2)^2 \right]
,
\label{Inv3}
\EEA
where $\psi^2={\Psi_i \Psi_i}$. 
Note that all terms entering (\ref{Inv3}) are non-negative and thus singularity in one of these terms
implies that the curvature invariant  $I_{{III}}$ is singular.
For non-vanishing expansion this is always the case for  $r=0$ and thus  a curvature singularity
is located there. For type III Kundt spacetimes, the invariant $I_{{III}}$ (and in fact all curvature invariants of all orders) 
identically vanishes \cite{Coleyetal04vsi}.

%------------------------------------------
\section{Type D and II }
\label{SecDII}

%----------------------------------------------------------

\subsection{Type D }
\label{sec_D}

In an adapted frame, type D Weyl tensor has only boost weight zero components determined by
$\WD{ij} $ and $C_{ijkl}$, see (\ref{Psi}), (\ref{Psi_SA}).

For vacuum type D spacetimes  with a parallelly propagated frame
and with the matrix $S_{ij}$  set to a diagonal form, Bianchi equations (\ref{B3}), (\ref{B5}) and (\ref{B12})  
%read
%\BEA
%-DR_{01ij}&=&R_{01ie}L_{ej}-R_{01je}L_{ei}+R_{0i1e}L_{ej}-R_{0j1e}L_{ei},\label{B3_D}\\
%\ \ DR_{0i1j}&=&-R_{0101}L_{ij}-R_{0i1e}L_{ej}-R_{01ie}L_{ej},\label{B5_D}\\
%-DR_{ijkm}&=&
%R_{0k1j}L_{im}+R_{0m1i}L_{jk}-R_{0k1i}L_{jm}\nonumber\\  &&
%-R_{0m1j}L_{ik}+R_{ijke}L_{em}-R_{ijme}L_{ek}.\label{B12_D}
%\EEA
%Since the matrix $S_{ij}$ is set to a diagonal form, in vacuum
%equations (\ref{B3_D}), (\ref{B5_D}) and (\ref{B12_D}) 
can be rewritten using (\ref{Psi}), (\ref{Psi_SA}),
cf also equations (24), (25) in \cite{PraPraOrt07}
\BEA
2D\WDA{ij} &=&-3\WDA{ij} (s_{(i)}+s_{(j)})-\WDS{ij} (s_{(j)}-s_{(i)}),\label{B3_D_Psi}\\
2D\WDS{ij} &=&3\WDA{ij} (s_{(i)}-s_{(j)})-\WDS{ij} (s_{(j)}+s_{(i)})-2\WD{} s_{(i)}\delta_{ij},\label{B5_D_Psi}\\
DC_{ijkm}&=&
-\WD{kj} s_{(i)}\delta_{im}-\WD{mi} s_{(j)}\delta_{jk}+\WD{ki} s_{(j)}\delta_{jm}
\nonumber\\ &&
+\WD{mj} s_{(i)}\delta_{ik}-C_{ijkm}(s_{(m)}+s_{(k)}).\label{B12_D_R}
\EEA
Equations (\ref{B3_D_Psi}) imply $\WDA{wv} =\Phi_{wv}^{A0}$.
For simplicity let us assume $\WDA{ij} =0$
and in what follows  we thus identify  $\WD{ij} $  with $\WDS{ij} $. Note that for Kerr-Schild spacetimes $A_{ij}=0$ $\Rightarrow$ $\WDA{ij} =0$ \cite{KSprep}, however, this
implication need not hold for general spacetimes. Then equations (\ref{B3_D_Psi}) yield
\BEA
\quad\quad\quad\quad\quad
\WD{pw} &=0,\label{Phi_pw}\\
\WD{pq} (s_{(q)}-s_{(p)})&=0, \ \ 
\label{B3_PS_pr}
\EEA
thus $\WD{pq} =0$ for $s_{(q)}\not= s_{(p)}$.

From equations (\ref{B5_D_Psi}), (\ref{B3_PS_pr}), for $p\not=q$ and  $s_{(q)}= s_{(p)}$
\BEA
2D\WD{wv} &=&0\ \ \Rightarrow\ \ \WD{wv} =\Phi_{wv}^0,\label{rce_PsiSww}\\
2D\WD{pq} &=&-\WD{pq} (s_{(p)}+s_{(q)})=-2\WD{pq} s_{(p)}\ \ \nonumber\\
& \Rightarrow&\ \ 
\WD{pq} =\frac{\Phi_{pq}^0}{r+a_{(p)}^0} \ \ \mbox{for}\ \ p\not=q,\ \ s_{(q)}= s_{(p)}.\label{PD_Psi_pr_ruzne}
\EEA

Trace of equations (\ref{B5_D_Psi}) together with (\ref{rce_PsiSww}) leads to
\BE
\fl
D\WD{} =D\WD{pp} 
 =-\WD{} S_{ii}-\WD{ii} s_{(i)}=-(\WD{pp} +\Phi_{ww}^0 )S_{ii}-\WD{pp} s_{(p)},
\label{rce_stopaPsi}
\EE
while the diagonal terms of (\ref{B5_D_Psi}) read
\BEA
D\WD{(p)(p)} &=&-(\WD{pp} +\Phi_{ww}^0 +\WD{(p)(p)} )s_{(p)}.\label{rce_PsiSpp}
\EEA

From now on we assume that  $s_{(p)}=1/r$ for all $p$ \footnote{In fact under this assumption   from equations (\ref{B3_D_Psi}) 
$\WDA{pq} =\Phi^{A0}_{pq}/r^3$, however, in what follows we still assume $\WDA{ij} =0$.}.
Then  equation (\ref{rce_stopaPsi}) reduces to
\BE
\fl
D\WD{} =D\WD{pp} =-\left(\WD{pp} +\Phi_{ww}^0\right){{\frac{m}{r}}}-\WD{pp} \frac{1}{r}
\ \ \Rightarrow\ \ \WD{pp} =\frac{\Phi^0}{r^{m+1}}-\frac{m\Phi_{ww}^0}{m+1}
\label{rce_stopaPsi_shearfree}
\EE
and thus
\BE
\WD{} =\frac{\Phi^0}{r^{m+1}}+\frac{\Phi_{ww}^0}{m+1}.\label{PD_stopaPsi}
\EE
Then equations (\ref{rce_PsiSpp}) imply
\BEA
 D\WD{(p)(p)}\!\!\!\!\!\!\!\!\!\!\!\! &=&-\left(\frac{\Phi^0}{r^{m+1}}+\frac{\Phi_{ww}^0}{m+1}  +\WD{(p)(p)} \right)s_{(p)}\nonumber\\
 \Rightarrow\  \WD{(p)(p)} &=&\frac{\Phi^0}{mr^{m+1}}+\frac{\Phi_{(p)(p)}^0}{r}-\frac{\Phi_{ww}^0}{m+1}. \label{PD_Psi_pp}
\EEA
Comparing  (\ref{PD_Psi_pp}) with (\ref{rce_stopaPsi_shearfree}) yields 
\BE
 \Phi_{pp}^0=0.\label{Phi_pp}
 \EE
Now we can combine (\ref{PD_Psi_pp}) with (\ref{PD_Psi_pr_ruzne}) in
\BE
\WD{pq} =\frac{\Phi_{pq}^0}{r} +\delta_{pq}\left(\frac{\Phi^0}{mr^{m+1}}-\frac{\Phi_{ww}^0}{m+1}\right).\label{Phi_pq}
\EE

From equations (\ref{B12_D_R}) for various combinations of indices we get
\BEA
\fl
C_{ij wv }&=&C_{ij wv}^0, %\ \ \   \Rightarrow \ \ \ R_{w_1 w_2 w_3 w_4}=R_{w_1 w_2 w_3 w_4}^0,
\label{D-R-1}\\
\fl
C_{ij wq }&=&\frac{C_{ij wq}^0}{r},\ \ \ {i,j\not=q},\label{D-R-2}\\
\fl
C_{p(q) w(q) }&=&\frac{C_{p(q) w(q)}^0}{r}, \label{D-R-3}\\
\fl
C_{v(q) w(q) }&=&\frac{C_{v(q) w(q)}^0}{r}+\Phi_{wv}^0,  \label{D-R-4}\\
\fl
C_{wvpq }&=&\frac{C_{wvpq}^0}{r^2},\label{D-R-5}\\
\fl
C_{wopq }&=&\frac{C_{wopq}^0}{r^2},\label{D-R-6}\\
\fl
C_{sopq}&=&-2(\delta_{sp}\delta_{oq}-\delta_{op}\delta_{sq})
\left( \Phi^0  \frac{F_m(r)}{r^2}
+\frac{\Phi_{ww}^0}{2(m+1)}\right)
\nonumber\\
\fl
&&+\frac{C_{sopq}^0}{r^2}+\frac{1}{r}(\Phi_{ps}^0\delta_{oq}+\Phi_{qo}^0\delta_{sp}-\Phi_{po}^0\delta_{sq}-\Phi_{qs}^0\delta_{op}),\label{D-R-7}
\EEA
where
\BE
\fl
F_m(r)=-\ln r 
\ \ \mbox{for}\ \ m=1,\ \ \ \ \ 
F_m(r)=\frac{1}{m(m-1)r^{m-1}}\ \ \mbox{for}\ \ m\not= 1.\label{Fm}
\EE
Note that some of the equations (\ref{D-R-1})--(\ref{D-R-7}) are not compatible with 
symmetries of the Weyl tensor  
unless corresponding  components vanish, thus
\BEA
C_{wpvz }&=&C_{vzwp }=0,\label{D-R-8}\\
C_{wvpq }&=&C_{pqwv }=0,\label{D-R-9}\\
C_{wopq }&=&C_{pqwo }=0 
\label{D-R-10}
\EEA
and from equations (\ref{eq:rcomps}) and (\ref{D-R-9}) 
\BE
C_{vpwq}^0=C_{vqwp}^0.\label{D-R-11}
\EE

Let us point out that for expanding type D  (and in general not for type II) spacetimes, Bianchi equations (\ref{B6}), 
  with $\Phi^0\not=0$, lead to
\BE
l_{1w}^0=0.\label{PD_l_1w}
\EE
However, we will not use this relation further in this section in order to obtain expressions also valid  for type II.

Using the identity $\WD{ij} =-{\textstyle{\frac{1}{2}}}C_{ikjk}$ (\ref{eq:rcomps_trace}) for the Weyl tensor we arrive at
\BEA
C_{vpwp}^0&=&0,\label{D-R-12}\\
C_{wzvz}^0&=&-(m+2)\Phi_{wv}^0,\\
C_{poqo}^0&=&0 \ \ \ \mbox{for}\ \ m\not=1,\label{PD_R_poqo=0}\\
\Phi^0&=&0\ \ \ \mbox{for}\ \ m=1,\\
C_{wpwq}^0&=&-m\Phi_{pq}^0.\label{R_Phi}
\EEA
Note that when $m=n-2$ (i.e. there are no  `$w$-type' indices), 
then $C_{pwqw}^0=0$ and thus from (\ref{R_Phi}) $\Phi^0_{pq}=0$.

To summarize: non-vanishing boost weight zero components of the Weyl tensor for type D (and II, see section \ref{sec_typeII}) vacuum spacetimes
with a non-twisting geodetic multiple WAND 
under the assumption $\WDA{ij} =0$ are $\WD{wv} $  and $\WD{pq} $ given in (\ref{rce_PsiSww}), (\ref{Phi_pq}), respectively,
\BEA
C_{vwyz}&=C_{vwyz}^0,\\
C_{vp wq }&=\frac{C_{vp wq}^0}{r}+\Phi_{wv}^0\delta_{pq},  
\EEA
 and $C_{sopq}$ given in (\ref{D-R-7}) with (\ref{Fm}), subject to (\ref{Phi_pp}), (\ref{D-R-11}), (\ref{D-R-12})--(\ref{R_Phi}).

From Ricci equations (\ref{11a})  
and (\ref{com_r_U}) with (\ref{PD_stopaPsi})
\BEA
\fl
L_{11}&=&- \left( l^0_{1w}l^0_{1w}+{\textstyle{\frac{1}{m+1}}}\Phi_{ww}^0 \right) r
+\frac{\Phi^0}{mr^{m}}+l_{11}^{0}, \label{L11D}\\
\fl
U&=&\left(
l^0_{1w} l^0_{1w}+{\textstyle{\frac{1}{2(m+1)}}}\Phi_{ww}^0 \right)r^2
-( l_{11}^{0}+ l_{1w}^0 \omega_ {w}^0)r
+\Phi^0 F_m(r)
+ U^0 \label{UD}
\EEA
and from (\ref{V}) 
\BE
\fl
V=-{\textstyle{\frac{1}{2(m+1)}}}\Phi_{ww}^0 r^2+r(l_{11}^{0}-l_{1w}^0\eta^{w0}_\alpha X^{\alpha 0})
-\Phi^0 F_m (r)
-U^0+X^{\alpha 0} 
\eta^{w0}_\alpha\omega^0_w.\label{PII-V}
\EE
 Then the metric component $g_{00}$  (\ref{metric00}) read
\BEA
\fl
g_{00}&=&\left(-{\textstyle{\frac{1}{(m+1)}}}\Phi_{ww}^0+\eta_\alpha^{p0}X^{\alpha 0}\eta_\beta^{p0}X^{\beta 0}+l_{1w}^0l_{1w}^0\right) r^2
+2r(l_{11}^{0}-2l_{1w}^0\eta^{w0}_\alpha X^{\alpha 0})\nonumber\\
\fl &&
-2\Phi^0 F_m(r) 
-2U^0+2X^{\alpha 0} 
\eta^{w0}_\alpha\omega^0_w
+\eta_\alpha^{w0}X^{\alpha 0}\eta_\beta^{w0}X^{\beta 0}\nonumber\\
\fl &=&
\left(\gamma^{2}-{\textstyle{\frac{1}{(m+1)}}}\Phi_{ww}^0\right)r^2+\gamma^{1}r+\gamma^{0}-2\Phi^0 F_m(r)
,\label{PD_g_00}
\EEA
where $\gamma^{N}$, $N=0,1,2$ are defined in (\ref{PIII_g00}).  
{\it The  metric for type D  vacuum spacetimes with a non-twisting geodetic multiple WAND then  has the form (\ref{typeIImetric}) with (\ref{PD_g_00}),
(\ref{IIg0i}), (\ref{cov.metric_D,III,N}) with $a_{(p)}^0=0$ and $l_{1w}^0=0$.} Note that (\ref{PD_g_00}) 
is valid  for type II spacetimes as well (see section \ref{sec_typeII}).

Let us now examine the Kretschmann scalar in vacuum
\BEA
R_{abcd}R^{abcd}&=&4R_{0101}^2+R_{ijkl}R_{ijkl}+8R_{0j1i}R_{0i1j}-4R_{01ij}R_{01ij}\nonumber\\
&=&4\WD{} ^2+C_{ijkl}C_{ijkl}+8\WDS{ij} \WDS{ij} -24\WDA{ij} \WDA{ij}.
\EEA
As was pointed out in \cite{KSprep}, under the assumption  $\WDA{ij} =0$, it reduces to
a sum of squares.  
Thus if any term $\Phi^0$, $\WD{pq} ^0$,  $C_{vpwq}^0$ or $C_{nopq}^0$ is non-zero,
then there is a scalar curvature singularity at $r=0$.

Note also that for asymptotically flat spacetimes 
 the Kretschmann scalar vanishes for $r\ \rightarrow\ \infty$ and thus in this case 
\BE
\Phi_{wv}^0=0=C_{wvyz}^0.
\EE

%---------------------------------------------------------

\subsection{Type II }
\label{sec_typeII}

Apart from boost weight zero components of the Weyl tensor, in type II spacetimes boost weight $-1$ components, $\Psi_i$,  
$\Psi_{ijk}$,
 and boost weight $-2$ components, $\Psi_{ij}$, also appear (see (\ref{Weyl_III}), (\ref{cykl})).
 However, these negative boost weight components do not enter Bianchi equations (\ref{B3_D_Psi})--(\ref{B12_D_R})
and thus assuming again $s_{(p)}=1/r$ for all $p$ and $\WDA{ij} =0$ all results obtained in  section \ref{sec_D}  for type D spacetimes except of (\ref{PD_l_1w}) are valid for type II spacetimes as well. 

In order to determine $r$-dependence of negative boost weight components of the Weyl tensor,
we analyse  the Bianchi equations (\ref{B1}), (\ref{B6}), (\ref{B9}) and (\ref{B4}), 
%\BEA
%\fl
%\quad\quad\quad\quad
%DR_{101i}-\delta_i R_{0101}&=-2R_{101k}L_{ki}-R_{0101}L_{i1}
%-R_{0i1k}L_{k1},
%\\ \fl
%\quad\quad\quad\quad\quad\quad\quad\quad
%-DR_{1kij}
%&=R_{1lij}L_{lk}+2R_{101[i}L_{j]k}
%+2R_{0k1[i}L_{j]1}
%-R_{klij}L_{l1}, \\ \fl
%\quad\quad\quad
%DR_{1i jk}+2\delta_{[k}R_{0|j]1i}&=2R_{1i[k|l}L_{l|j]}+2R_{0[k|1l}\Ml_{i|j]}-2R_{0l1i}\Ml_{[jk]},
%\\
%\fl
%DR_{1i1j}-\Delta R_{0j1i}-\delta_j R_{101i}&=-R_{1i1l}L_{lj}
%+R_{1ijl}L_{l1}+R_{101l}\Ml_{ij}+R_{0101}N_{ij}
%\nonumber \\  
%& \quad
%+R_{0l1i}N_{lj}
%+R_{0j1l}\Ml_{i1}+R_{0l1i}\Ml_{j1},
%\EEA
which can be rewritten as
\BEA
\fl
\quad\quad\quad\quad\quad\quad
D\Psi_{i}-\delta_i \Phi&=-2\Psi_{i}s_{(i)}-\Phi L_{i1}
-\WDS{ik} L_{k1},\label{II-B1}\\
\fl
\quad\quad\quad\quad\quad\quad\quad
-2D\Psi_{ijk}
&=2\Psi_{ijk}s_{(k)}+2\Psi_{[i}\delta_{j]k}s_{(k)}
+2\WDS{k[i} L_{j]1}
-C_{klij}L_{l1},
\label{II-B6}\\
\fl
\quad\quad\quad
2D\Psi_{ jki}+2\delta_{[k}\WDS{j]i} & =2\Psi_{kji}s_{(j)}
-2\Psi_{jki}s_{(k)}
+2\WDS{[k|l} \Ml_{i|j]}-2\WDS{li} \Ml_{[jk]},\label{II-B9}\\
\fl
\quad
2D\Psi_{ij}-\Delta \WDS{ij} -\delta_j \Psi_{i}&=-2\Psi_{ij}s_{(j)}
+2\Psi_{jli}L_{l1}+\Psi_{l}\Ml_{ij}
+\WD{} N_{ij}
\nonumber\\  \ \ \ \ \ \ \ \ \ \ \ 
&\ \ +\WDS{li} N_{lj}
+\WDS{jl} \Ml_{i1}+\WDS{li} \Ml_{j1}.\label{II-B4}
\EEA

Using previous results,
from (\ref{II-B1}) 
\BEA
\fl
\Psi_{w}&=&r\left[{\textstyle{\frac{1}{m+1}}}(\xi^{\alpha 0}_w\Phi_{vv}^0,_\alpha -l_{1w}^0\Phi_{vv}^0)-\Phi_{wv}^0 l_{1v}^0\right]+\Psi_w^0
\nonumber\\ \fl &&
-(l_{1w}^0\Phi^0m+\xi^{\alpha 0}_w\Phi^0,_\alpha){\textstyle{\frac{1}{mr^m}}}
+{\textstyle{\frac{1}{r^{m+1}}}}\omega^0_w\Phi^0,\label{PII_Psi_w}\\ 
\fl
\Psi_{p}&=&{\textstyle{\frac{1}{2(m+1)}}}\xi^{\alpha 0}_p\Phi_{ww}^0,_\alpha+{\textstyle{\frac{1}{r^2}}}\Psi^0_p
-\xi^{\alpha 0}_p\Phi^0,_\alpha {\textstyle{\frac{mF_m(r)}{r^2}}}\label{PII_Psi_p}
\EEA
and from (\ref{II-B6}) 
\BEA
\fl
\Psi_{pqw}&=&\Psi_{pqw}^0,\\
\fl
\Psi_{vpw}&=&\Psi_{vpw}^0,\\
\fl
\Psi_{vzw}&=&\pul (\Phi_{wz}^0l_{1v}^0-\Phi_{wv}^0l_{1z}^0+C_{wyvz}^0l_{1y}^0) r+\Psi_{vzw}^0,\\
\fl
\Psi_{vwp}&=&\textstyle{\frac{1}{r}}{\Psi_{vwp}^0},\\
\fl
\Psi_{wpq}&=&
-\textstyle{\frac{r}{4(m+1)}} \xi^{\alpha 0}_w\Phi_{vv}^0,_\alpha \delta_{pq}
+\pul(\Phi_{pq}^0l_{1w}^0+C_{qzwp}^0l_{1z}^0-\Psi_w^0\delta_{pq})
+\textstyle{\frac{1}{r}}{\Psi_{wpq}^0}\nonumber\\ \fl &&
-{\textstyle{\frac{\delta_{pq}}{2}}} [l_{1w}^0\Phi^0(m+1)+\xi^{\alpha 0}_w\Phi^0,_\alpha ]\textstyle{\frac{F_m(r)}{r}}
+\textstyle{\frac{\delta_{pq}}{2mr^{m+1}}}\omega_w^0\Phi^0,\\
\fl
\Psi_{oqp}&=&\!
\textstyle{\frac{1}{4(m+1)}}\Phi_{ww}^0,_\alpha (\delta_{op}\xi^{\alpha 0}_q-\delta_{pq}\xi^{\alpha 0}_o)
+\Psi_{oqp}^0\textstyle{\frac{1}{r}}
+\textstyle{\frac{1}{2r^2}}(\delta_{pq}\Psi^0_o-\delta_{po}\Psi^0_q)\nonumber\\ \fl &&
+\Phi^0,_\alpha \textstyle{\frac{F_m(r)}{2r^{2}}}(\delta_{po}\xi^{\alpha 0}_q-\delta_{pq}\xi^{\alpha 0}_o).\label{PII_Psi_oqp}
\EEA
The Weyl components
$\Psi_i$ and $\Psi_{ijk}$ as given in (\ref{PII_Psi_w})--(\ref{PII_Psi_oqp}) are subject to (\ref{cykl})
and therefore 
\BEA
\Psi_{wvp}^0&=&0,\ \ \  \Psi_{pqw}^0=0,\ \ \  \Psi_{pwv}^0=\Psi_{pvw}^0,\ \ \
\Psi_{wpq}^0=\Psi_{wqp}^0,\\
0&=&\Psi_{vwz}^0+\Psi_{zvw}^0+\Psi_{wzv}^0,\\
0&=&\Psi_{pqo}^0+\Psi_{opq}^0+\Psi_{qop}^0,\\
C_{vywv}^0l_{1y}^0\!\!\!\!\!\!\!\!\!&=&\textstyle{\frac{m+2}{2(m+1)}}\xi^{\alpha 0}_w\Phi^0_{zz},_\alpha
-\textstyle{\frac{m+2}{(m+1)}}l_{1w}^0\Phi_{zz}^0,\\
2\Psi_{wzz}^0&=&l_{1z}^0 C_{zpwp}^0+\Psi_w^0(m+1),\\
\Psi_{wpp}^0&=&\pul l_{1w}^0\Phi^0 \ \ \mbox{and}\ \ \ 2l_{1w}^0\Phi^0+\xi^{\alpha 0}_w\Phi^0,_\alpha=0\ \  \mbox{for}\ \ \ m=1,\\
\Psi_{wpp}^0&=&0\ \ \ \mbox{and}\ \ \ 2ml_{1w}^0\Phi^0+\xi^{\alpha 0}_w\Phi^0,_\alpha=0\ \ \ \mbox{for}\ \ \ m>1,\\
\Psi_{pww}^0&=&\textstyle{\frac{m}{4(m+1)}}\xi^{\alpha 0}_p\Phi^0_{ww},_\alpha ,\\
\Psi_{pqq}^0&=&0,\\
\Psi_p^0&=&0\ \ \mbox{and}\ \ \ \xi^{\alpha 0}_p\Phi^0,_\alpha =0\ \ \ \mbox{for}\ \ \ m=1,\\
\Psi_p^0(m-2)\!\!&=&0\ \ \mbox{and}\ \ \ (m-2)\xi^{\alpha 0}_p\Phi^0,_\alpha =0\ \ \ \mbox{for}\ \ \ m>1.\label{PII_Psi_p,Phi0}
\EEA

In order to determine $r$-dependence of $\Psi_{ij}$ from  equations (\ref{II-B4}), first  we need to 
find $\Mi_{j1}$ and $N_{ij}$. Note that for $\WDA{ij} =0$, 
the Ricci equations  (\ref{11m}) reduce to those of the Weyl type III  with solution given in (\ref{PIII_Mij1}).
From Ricci equations  (\ref{11j}), %$DN_{ij}=-N_{ik}L_{kj}-C_{0j1i}$,
\BEA
\fl
N_{pw}&=&n_{pw}^0, \ \ \ N_{vw}=-\Phi_{wv}^0r+n_{vw}^0,\ \ \ N_{wp}=\textstyle{\frac{n_{wp}^0}{r}},\ \ \\ 
\fl
N_{pq}&=&-\Phi_{pq}^0+\textstyle{\frac{n_{pq}^0}{r}}
+\delta_{pq}\left[ r\textstyle{\frac{\Phi_{ww}^0}{2(m+1)}}
+\Phi^0\textstyle{\frac{F_m(r)}{r}}\right].
\EEA
Now $r$-dependence of $\Psi_{ij}$ can be determined from equations (\ref{II-B4})
\BEA
\fl
\Psi_{vw}&=&r^2\Psi_{vw}^{A}+r\Psi_{vw}^B+\Psi_{vw}^C 
-\textstyle{\frac{\mpo{vw} \Psi_p^0}{2r}}
-\mpo{vw} \xi^{\alpha 0}_p\Phi^0,_\alpha \textstyle{\frac{\ln r+1}{2r}}\delta_{1m}\nonumber\\ \fl &&
+mF_m(r)\Psi_{vw}^D
+\textstyle{\frac{1}{r^m}}\Psi_{vw}^E+\textstyle{\frac{1}{r^{m+1}}}\Psi_{vw}^F,\\
\fl
\Psi_{pw}&=&r^2\Psi_{pw}^A+r\Psi_{pw}^B+
\Psi_{pw}^C\ln r
+\Psi_{pw}^D
+\textstyle{\frac{\ln r}{r}}\delta_{1m}\Psi_{pw}^E
+\textstyle{\frac{1}{r}}\Psi_{pw}^F
\nonumber\\ \fl &&
+\textstyle{\frac{\ln r}{r^2}}\delta_{1m}\Psi_{pw}^G
+\textstyle{\frac{1}{r^2}}\Psi_{pw}^H
+\textstyle{\frac{1}{r^{m-1}}}\Psi_{pw}^I
+\textstyle{\frac{1}{r^m}}\Psi_{pw}^J+\textstyle{\frac{1}{r^{m+1}}}\Psi_{pw}^K,
\\ \fl
\Psi_{pq}&=&r^2\Psi_{pq}^A\delta_{pq}+r\Psi_{pq}^B
+\ln r\delta_{1m}\Psi_{pq}^C+\Psi_{pq}^D
+\textstyle{\frac{\ln r}{r}}\Psi_{pq}^E
+\Psi_{pw}^F
\textstyle{\frac{1}{r}}
+\textstyle{\frac{\ln r}{r^2}}\delta_{1m}\Psi_{pq}^G\nonumber\\ \fl &&\!\!\!\!\!
-\textstyle{\frac{1}{r^2}}\Psi_{pq}^H 
+\textstyle{\frac{1}{r^{m-1}}}\Psi_{pq}^I
+\textstyle{\frac{1}{r^{m}}}\Psi_{pq}^J
+\textstyle{\frac{1}{r^{m+1}}}\Psi_{pq}^K,
\EEA
where $\Psi_{ij}^A$, $\Psi_{ij}^B$, $\dots$ $\Psi_{ij}^K$ do not depend on $r$.
Since in this paper we are mainly interested in the $r$-dependence of the metric and the Weyl tensor we do not
give here quite complicated explicit expressions for $\Psi_{ij}^A$, $\Psi_{ij}^B$, $\dots$  $\Psi_{ij}^K$.

%----------------------------------------------------
\subsubsection{The case with $L_{1i}=0$.}
\label{sec_spec-vanishL1i}

When (\ref{PD_l_1w}) is satisfied (for type D and special cases of other Weyl types considered here) then
$\omega_w$ can be transformed away by  null rotation with fixed $\bl$ (\ref{tr-nullrot}) %\cite{Milsonetal05,OrtPraPra07} 
with $z_w=\omega_w^0$ (\ref{rot_om}) and thus (assuming all $s_{(p)}$ are the same) $\omega_{i}=0$ for all $i$. 
Since now $g^{1\alpha}=X^{\alpha 0}$,     
we introduce $\tilde x^\alpha=\tilde x^\alpha (x^\beta,u)$ as in
\cite{PodOrt06},  leaving unchanged  null hypersurfaces
$u=$const and preserving the affine character of the parameter $r$,  
to set $\tilde g^{1\alpha}=0$, i.e. (omitting the tilde symbol) 
\BE
X^{\alpha 0}=0 . 
\EE
Then from (\ref{YN})--(\ref{V}) and (\ref{PII-V}) 
we get 
\BE
\fl
V=-U=-{\textstyle{\frac{1}{2(m+1)}}}\Phi_{ww}^0 r^2+rl_{11}^{0}
-\Phi^0 F_m (r)
-U^0,\ \ \Omega^{i}=0,\ \ Y_\alpha=0.\label{V_vanishL1i}
\EE

Equations (\ref{metric11})--(\ref{cov.metric_D,III,N})
now reduce to
\BEA
\fl
g_{11}&=&0,\ \ \  g_{01}=1,\ \ \ g_{1\alpha}=0,\ \ \ g_{00}=2V, \ \ \ 
g_{0\alpha}=0,\label{IIgS00}\\
\fl
g_{\alpha\beta}&=&\eta_\alpha^k\eta_\beta^k=\eta_\alpha^{p0}\eta_\beta^{p0}r^2+\eta_\alpha^{w0}\eta_\beta^{w0}
=\gamma_{\alpha\beta}^{2}r^2+\gamma_{\alpha\beta}^{0}
,\label{cov.metricS_D,III,N}
\EEA
and thus {\it the metric of vacuum spacetimes with a non-twisting  geodetic multiple WAND (i.e. types II, D, III or N) with $L_{1i}=0$  can be set into the form}
%reads
\BE
\fl
{\mbox{d}}s^2=2V{\mbox{d}}u^2+2{\mbox{d}}u{\mbox{d}}r+\left(\gamma_{\alpha\beta}^{2}r^2+\gamma_{\alpha\beta}^{0}\right)\!{\mbox{d}}x^\alpha{\mbox{d}}x^\beta, \ \ \ \ 
\label{typeIImetric_vanishL1i}
\EE
{\it where functions $\gamma_{\alpha\beta}^{N}$, $N=0,2$, introduced in  (\ref{cov.metricS_D,III,N}) do not depend on $r$
and $V$ is given in (\ref{V_vanishL1i}). }

%-------------------------------------------------------------------------------
\subsubsection{The shearfree case}

Let us now briefly discuss the shear-free case which occurs for $m=0$ (Kundt spacetimes) and for $m=n-2$ (Robinson-Trautman spacetimes \cite{PodOrt06}).

Kundt spacetimes in vacuum are necessarily of type II or more special  \cite{OrtPraPra07} and they thus form $m=0$ subclass of spacetimes studied in the present paper. Note that in contrast to the expanding case, the components of the metric (\ref{typeIImetric}), including $g_{00}$, are at most quadratic polynomials in $r$. Similarly as in four dimensions boost weight 0, -1 and -2 components of the Weyl tensor are independent on $r$, linear and quadratic in $r$, respectively.

 In the $m=n-2$ case in four dimensions,
 equations (\ref{PII_Psi_p,Phi0})  are identically satisfied
 and consequently the corresponding  class of Robinson-Trautman spacetimes is very rich and includes e.g.  radiative type N and III spacetimes
 as well as type D C-metric describing uniformly accelerated black holes emitting gravitational radiation.
 However, in higher dimensions equations (\ref{PII_Psi_p,Phi0})   imply
$\Psi_p^0=0$ and using (\ref{norm_delta}) $\Phi^0,_\alpha=0$. 
 From (\ref{R_Phi}) $\Phi^0_{pq}=0$
and then from (\ref{11l}) or (\ref{II-B9}) we  get $\Psi_{oqp}^0=0$.
Therefore all components of the Weyl tensor
with boost weight $-1$ vanish. Similarly it can be shown that boost weight $-2$ components of the Weyl tensor
vanish as well. 
Thus in higher dimensions vacuum shear-free spacetimes admitting non-twisting geodetic multiple WAND
  are necessarily of type D in agreement with \cite{PodOrt06}.
The Weyl tensor is now given by 
\BE
\fl
\WD{pq} =\delta_{pq}\frac{\Phi^0}{(n-2)r^{n-1}},\ \ \ 
C_{sopq}=-2(\delta_{sp}\delta_{oq}-\delta_{op}\delta_{sq})  \Phi^0  \frac{F_{(n-2)}(r)}{r^2}+\frac{C_{sopq}^0}{r^2}.
\EE
Note that in four dimensions equation (\ref{PD_R_poqo=0}) implies $C_{sopq}^0=0$, while in higher dimensions this
term, corresponding essentially to the curvature of the spatial part of the metric $\gamma_{\alpha\beta}^2$ \cite{PodOrt06}, in general does not vanish. Therefore the $r$-dependence of the Weyl tensor and thus also the asymptotic behaviour of gravitational field in higher dimensions is more complex than in four dimensions\footnote{Note  that in boost weight zero Weyl components in the $m<n-2$ case  terms proportional to $r^0$ and $r^{-1}$ also appear.}.
This is, however, beyond the scope of the present paper and will be studied elsewhere.
  
%-----------------------------------------------------------------

\section{Construction of 
an explicit  expanding  type N solution  in five dimensions with $l_{14}^0=0$}
\label{Sec5D}

Apart from usual motivation coming from higher-dimensional general relativity, there is an additional reason for studying type N vacuum spacetimes.
For these spacetimes all curvature invariants involving metric, the Riemann  tensor and its first covariant derivatives vanish. Such solutions thus belong
to ${\rm VSI}_1$ class of spacetimes \cite{VSIi}, which are solutions of various field theories to all
orders with a specific effective action containing only certain higher order correction terms (see \cite{VSIi}).

Let us explicitly mention the Einstein-Gauss-Bonnet equations
\BEA
%G_{ab} 
\fl
R_{ab} - \pul Rg_{ab} = %-\Lambda g_{ab}+
          \alpha\left(\pul\mathcal{L}_{GB}g_{ab} - 2RR_{ab}
         + 4R_{ac}R_{b}^{\phantom{b}c} + 4R_{acbd}R^{cd}
         - 2R_{acde}R_{b}^{\ cde}\right), \label{EGB}
%G_{ab} &\equiv& R_{ab} - \pulRg_{ab} \; .\nonumber
\EEA
where $\mathcal{L}_{GB} = R^{2} - 4R_{ab}R^{ab} + R_{abcd}R^{abcd}$ and $\alpha$  is the Gauss-Bonnet coupling constant.
It can be seen directly  that vacuum type N solutions to the Einstein equations solve vacuum  
Einstein-Gauss-Bonnet equations (\ref{EGB}) as well since for these spacetimes $R_{acde}R_{b}^{\ cde}=0=R_{abcd}R^{abcd}$.

In this section,  we  attempt to 
derive an expanding non-twisting type N vacuum solution and we limit ourselves  to a five-dimensional case with an additional assumption $l_{14}^0=0$.
Since  resulting metrics we have obtained so far can be obtained
by taking a direct product of  four-dimensional type N vacuum metrics with an extra dimension,  the main purpose of this section is thus to illustrate 
the use of the higher-dimensional NP formalism for constructing exact vacuum solutions. 
Note that corresponding Bianchi and Ricci  equations are quite complex and thus at several points of the calculation we make various assumptions in order to simplify them. This, however, obviously comes with the price of possibly reducing the resulting class of solutions.

%----------------------------------------------------------------

For 
explicit calculations it turns out to be more convenient to 
relax the assumption of diagonal $\Psi_{ij}$ from section \ref{sec_WeylN}
and so now there are  two independent components of the Weyl tensor
 $\Psi_{33}=-\Psi_{22}$, $\Psi_{32}=\Psi_{23}$ with the rest of the 
components vanishing. Therefore we cannot use the form of the Weyl tensor
obtained in (\ref{pN}) and instead from
the Bianchi equations (\ref{B4}) %=23 in \cite{Pravdaetal04}  
\BE
\Psi_{22}=-\Psi_ {33}=\frac{p^0}{r}, \ \ \ \Psi_{23}=\Psi_ {32}=\frac{\Pi^0}{r},\ \ \ \ \Psi_{pw}=0=\Psi_{wv}.\label{pN5d}
\EE

Assuming $l_{14}^0=0$, NP equations simplify considerably and are given in  \ref{appN5nulL1i}.
In fact the following quantities vanish:
\BEA
\mDo{43} =\mTo{42} =\mDo{44} =\mTo{44} 
%=\mDo{34}
 =0\label{PN-zero_m},\\
n_{24}^0=n_{34}^0=\mDo{41} =\mTo{41} =n_{23}^0=n_{32}^0={   \omega}_2^0={   \omega}_3^0=0.\label{PN-zero_n}
\EEA

Similarly as in section \ref{sec_spec-vanishL1i} we transform away $\omega_4^0$.
{However,  here we  do not transform away the functions $X^{\alpha 0}$.}
Then from equations (\ref{Ricci_k_2s}) 
\BE
{    \omega}_4^0=\mCo{22} =\mCo{33} =0\label{Ricci_k_2ss},
\EE
and from (\ref{Ricci_p_m422_2s})--(\ref{Ricci_p_m422_4s}) and (\ref{Ricci_o_4s})
\BE
n_{42}^0=n_{43}^0=n_{44}^0=n_{41}^0=0.\label{n42=0}
\EE
From (\ref{Ricci_i_Us})   
 we get $U^0=n_{22}^0=n_{33}^0$
and then equations (\ref{com_r_delU_1s})--(\ref{com_r_delU_3s})
(now identical with (\ref{Ricci_l_1s})--(\ref{Ricci_l_3s})) imply  
\BE
U^0=n_{22}^0=U^0(u).
\EE
Let us assume $U^0=n_{22}^0=$const.

Apart from $l_{14}^0=0$ we make the following simplifying assumptions:
\BEA
\mDo{34}\!\!\! &=&0,\label{m234_0}\\
\ \ \xi^3_3&=&-\xi^2_2\not=0, \ \ \ 
\xi^4_4\not=0,\ \ \ \ \mbox{all other}\ \xi^\alpha_k=0.\label{assump_diag_xi}
\EEA
Note that $\mDo{34} $ always vanishes for diagonal $\Psi_{ij} $,  see (\ref{Bianchi_B10_3bS}).

Under the assumptions (\ref{assump_diag_xi}) 
from  (\ref{com_x_delxi_2s}), (\ref{com_x_delxi_3s}) (\ref{Ricci_c_l11,4s}),  (\ref{Ricci_h_3S}), (\ref{Ricci_h_6S}),
(\ref{Ricci_p_1s}), (\ref{Ricci_p_2s}), (\ref{Ricci_o_3s}), 
%(\ref{com_x_delxi_2s})--(\ref{com_x_delX_2s}),
(\ref{com_x_delX_3s}),
(\ref{Bianchi_B10_3S}), (\ref{Bianchi_B10_3bS}) we obtain that $\xi^{20}_2=-\xi^{30}_3$, $l_{11}^0$, $n_{21}^0$, $n_{31}^0$,
$\mTo{22} \!$, $\mDo{33} \!$, $\mDo{31} \!$, 
$X^{20}$, $X^{30}$, $p^0$, $\Pi^0$ do not depend
on $x^3=z$. From (\ref{com_x_delxi_2s})--(\ref{com_x_delX_2s})
it also follows that $\xi^{40}_4=\xi^{40}_4(u,z)$, $X^{40}=X^{40}(u,z)$ are functions of $u,z$ only.
 
Equations (\ref{com_x_delxi_1s}), (\ref{Ricci_p_3s})
can be rewritten using (\ref{assump_diag_xi}) 
\BEA
\quad \quad\quad\quad\quad\quad\quad\quad\quad \quad\quad\quad\quad\quad\quad
{\xi^{20}_{2}},_3&=&
\mTo{22}  ,\label{com_x_delxi_1AS}\\
\quad \quad\quad\quad\quad\quad\quad\quad\quad \quad\quad\quad\quad\quad\quad
 {\xi^{20}_{2}},_2&=&-\mDo{33}  ,\label{com_x_delxi_1BS}\\
{\xi^{20}_{2}}({\xi^{20}_{2}},_{22}+{\xi^{20}_{2}},_{33}) -({\xi^{20}_{2}},_2)^2-({\xi^{20}_{2}},_3)^2&=&
2n_{22}^0.
\label{Ricci_p_x22S}
\EEA
Assuming $\xi^{20}_2$ to have a form
of a polynomial in $x^2=x$ and $x^3=y$,
 after an appropriate translation in $x$, $y$,
we arrive at
\BE
\fl
\xi^{20}_2\!=\!A_0P(x,y), \ \   P(x,y)\!=\!(1+ex^2+ey^2),\ \  \mDo{33} =-2A_0ex,\  \ \mTo{22} =2A_0ey, \ \ \
\EE
where we set $A_0=1/\sqrt{2}$ and $e=n_{22}$ is assumed to be independent on $u$.
From  (\ref{com_x_delX_1s}) and  (\ref{com_x_delX_2s}) it follows
\BE
X^{20},_2=X^{30},_3,\ \ \  \ X^{30},_2=-X^{20},_3,
\EE
with the integrability condition $X^{20},_{22}+X^{20},_{33}=0$
and from (\ref{com_x_delX_1s}) 
\BE
l_{11}^0=-\frac{2e(xX^{20}+yX^{30})}{P(x,y)}+X^{20},_x.
\EE
Then (\ref{Ricci_c_l11,2s}), (\ref{Ricci_c_l11,3s}) determine $n_{21}^0$, $n_{31}^0$
\BEA
\fl
n_{21}&=-\textstyle{\frac{\sqrt{2}}{2}}P(x,y)X^{20},_{xx}
+\sqrt{2}e(X^{20}+xX^{20},_x-yX^{20},_y)-2\sqrt{2}ex\textstyle{\frac{xX^{20}+yX^{30}}{P(x,y)}},\\
\fl
n_{31}&=\ \textstyle{\frac{\sqrt{2}}{2}}P(x,y)X^{20},_{xy}
-\sqrt{2}e(X^{30}+xX^{20},_y+yX^{20},_x)+2\sqrt{2}ey\textstyle{\frac{xX^{20}+yX^{30}}{P(x,y)}}.
\EEA
From equation (\ref{Ricci_h_n22u_n21S}) or (\ref{Ricci_h_n22u_n31S}) and from (\ref{Ricci_h_2S}) or (\ref{Ricci_h_4S}) we get 
\BEA
p^0&=-&\pul X^{20},_{xxx}P(x,y)^2,\\  %\pul X^{20},_{xyy}P(x,y)^2
\Pi^0&= &\pul X^{20},_{xxy}P(x,y)^2.   %-\pul X^{20},_{yyy}P(x,y)^2
\EEA

Equations (\ref{YN})--(\ref{V}) 
 lead to $V=-U$, $Y_j=0$, $\Omega^i=-\eta^{(i)}_{(i)}X^{(i)0}=-X^{(i)}/\xi^{(i)}_{(i)}$. 
The contravariant frame vectors now read
\BEA
\fl
\ell^a&=&[0,1,0,0,0],\\
\fl
n^a&=&
\left[ 1,-\left( -\textstyle{\frac{2e(xX^{20}+yX^{30})}{P(x,y)}}+X^{20},_x\right) r+e,X^{20},X^{30},X^{40}\right], \\
\fl
m^a_{(2)}&=&
A_0P(x,y){\textstyle{\frac{1}{r}}}[0,0,1,0,0],\\
\fl
m^a_{(3)}&=&
-A_0P(x,y){\textstyle{\frac{1}{r}}}[0,0,0,1,0],\\
\fl
m^a_{(4)}&=&
\xi^{40}_4[0,0,0,0,1];
\EEA
and the covariant frame vectors are
\BEA
\fl
\ell_a&=&[1,0,0,0,0],\\
\fl
n_a&=&
\left[ \left( -\textstyle{\frac{2e(xX^{20}+yX^{30})}{P(x,y)}}+X^{20},_x\right) r-e,1,0,0,0\right],\\
\fl
m_a^{(2)}&=&
\textstyle{\frac{r}{A_0P(x,y)}}[-X^{20},0,1,0,0],\\
\fl
m_a^{(3)}&=&
-\textstyle{\frac{r}{A_0P(x,y)}}[-X^{30},0,0,1,0],\\
\fl
m_a^{(4)}&=&
\left[-\textstyle{\frac{X^{40}}{\xi^{40}_{4}}},0,0,0,\textstyle{\frac{1}{\xi^{40}_4}}\right],
\EEA
where $\xi^{40}_{4}$, $X^{40}$ are subject to (\ref{com_x_delX_3s}), i.e.
\BE
-{\xi^{40}_{4}},_u- X^{40}{\xi^{40}_{4}},_4+ {\xi^{40}_{4}} X^{40},_4= 
0. 
\label{com_x_delX_3S}
\EE

The metric thus reads
\BEA
\fl
{\mbox{d}}s^2=
\left[ 2l_{11}^0r-2e\! +\!\left(\textstyle{\frac{r}{\xi^{20}_2}}\right)^2\!\left((X^{20})^2\!+\!(X^{30})^2\right)
+\left(\textstyle{\frac{X^{40}}{\xi^{40}_4}}\right)^2\right]{\mbox{d}}u^2
\!+\!2{\mbox{d}}u{\mbox{d}}r  \nonumber\\  \fl 
\!-\!2{\mbox{d}}u\!\left[ \left( \textstyle{\frac{r}{\xi^{20}_2}}\right)^2\!(X^{20}{\mbox{d}}x\!+\!X^{30}{\mbox{d}}y)
\!+\!\left(\textstyle{\frac{1}{\xi^{40}_4}}\right)^2\!X^{40}{\mbox{d}}z\right]\!
+\!\left(\textstyle{\frac{r}{\xi^{20}_2}}\right)^2\! ({\mbox{d}}x^2\!+\!{\mbox{d}}y^2) \!+\!\left(\textstyle{\frac{1}{\xi^{40}_4}}\right)^2\!\!{\mbox{d}}z^2\!.
\label{spec5dmetric}
\EEA
Introducing  $\tilde z=\int 1/\xi^{40}_4 {\mbox{d}}z$ and using (\ref{com_x_delX_3S})
 the metric (\ref{spec5dmetric}) reduces to 
\BEA
\fl
{\mbox{d}}s^2&=&
\left[ 2l_{11}^0r-2e +\left(\textstyle{\frac{r}{\xi^{20}_2}}\right)^2\left((X^{20})^2+(X^{30})^2\right)\right]
{\mbox{d}}u^2+2{\mbox{d}}u{\mbox{d}}r\nonumber\\
\fl
&&
-2{\mbox{d}}u \left( \textstyle{\frac{r}{\xi^{20}_2}}\right)^2(X^{20}{\mbox{d}}x+X^{30}{\mbox{d}}y)
+\left(\textstyle{\frac{r}{\xi^{20}_2}}\right)^2 ({\mbox{d}}x^2+{\mbox{d}}y^2) +{\mbox{d}}\tilde z^2, \label{RTN-DP}
\EEA
where 
\BE
{\mbox{d}}\tilde z
=\textstyle{\frac{1}{\xi^{40}_4}}{\mbox{d}}z
-\textstyle{\frac{X^{40}}{\xi^{40}_4}}{\mbox{d}}u.
\EE
So the metric (\ref{RTN-DP}) represents a direct product of a four-dimensional Robinson-Trautman type N vacuum solution
with an extra dimension.

%---------------------------------------------------------

\ack
V.P. and A.P. acknowledge support from research plan No
AV0Z10190503 and research grant KJB100190702. We are grateful to Marcello Ortaggio for reading the manuscript and for discussions concerning \cite{PodOrt06} and to Harvey Reall for useful comments.

\appendix

%--------------------------------------------------------

\section{Ricci identities}

\label{sec_ricci}

The Ricci equations, i.e. contractions of the Ricci identities $v_{a;bc}-v_{a;cb}={R}_{sabc}v^s$ with the frame vectors
(\ref{normalization}), 
in higher dimensions, given in full generality in equations (11a)--(11p) in \cite{OrtPraPra07}, are rewritten here for vacuum spacetimes with a geodetic multiple WAND
(Weyl types II, D, III or N)
in a parallelly propagated frame
%\numparts
\BEA
\fl 
  DL_{11} &=  - L_{1i} L_{i1} 
%\nonumber \label{Lnln} \\
 %\qquad\qquad\qquad {}
- C_{0101} 
%+ \textstyle{\frac{2}{n-2}} R_{01} - \textstyle{\frac{1}{(n-1)(n-2)}}R
\, ,  \label{11a}\\
%--------------------------------------------------------
\fl
DL_{1i}& = 
-L_{1j} L_{ji} % +  \textstyle{\frac{1}{n-2}}R_{0i}
\, , \label{11b}\\
%--------------------------------------------------------
\fl
   \bigtriangleup L_{1i} - \delta_i L_{11}& =  
 L_{11}(L_{1i}- L_{i1}) - 2 L_{j[1|} N_{j|i]} 
 %\nonumber\\ 
 %\qquad\qquad\qquad 
 %&{}
 -L_{1j} (N_{ji}+\Mj_{{i}{1}}) + C_{101i} %- \textstyle{\frac{1}{n-2}} R_{1i}
\, , \label{11c}\\
%--------------------------------------------------------
\fl
\delta_{[j|} L_{1|i]}& = - L_{11} L_{[ij]}  
-  L_{1k} \Mk_{{[i}{j]}} 
%\nonumber\\ 
% \qquad\qquad\qquad 
%&{}
- L_{k[j|} N_{k|i]} + \textstyle{\frac{1}{2}}C_{01ij}\, ,  \label{11d}\\
%--------------------------------------------------------
\fl
  D L_{i1}& =   
- L_{ij} L_{j1}  % + \textstyle{\frac{1}{n-2}}R_{0i}
\, , \label{11e}\\
%--------------------------------------------------------
\fl
   - D N_{i1}& =    
 N_{ij}  L_{j1} - C_{101i} %+ \textstyle{\frac{1}{n-2}}R_{1i}
\, , \label{11f}\\
%--------------------------------------------------------
\fl
D L_{ij} &=  
 - L_{ik} L_{kj} % -\textstyle{\frac{1}{n-2}}R_{00} \delta_{ij}
\, ,  \label{11g} \\%\label{Lmlm}\\
 %--------------------------------------------------------
\fl
\bigtriangleup N_{ij} - \delta_j N_{i1} &= - L_{11} N_{ij}  
- N_{i1} (-2 L_{1j} + L_{j1})   
\nonumber\\
% \qquad\quad 
&
{}+ 2 N_{k[1|} \Mk_{{i|}{j]}} - N_{ik} (N_{kj} + \Mk_{{j}{1}})   
 - C_{1i1j} %- \textstyle{\frac{1}{n-2}} R_{11}\delta_{ij}
\, ,  \label{11h}\\
  %--------------------------------------------------------
\fl
\bigtriangleup L_{ij} - \delta_j L_{i1}& =  L_{11} L_{ij}  
- L_{i1} L_{j1} + 2 L_{k[1|} \Mk_{{i|}{j]}} 
%\nonumber\\
%  \quad \!\!\!\!\!\!\!\! \!\!\!\!\!\!\! 
%&{}
- L_{ik} (N_{kj} + \Mk_{{j}{1}})   
 - C_{0i1j} 
%-  \textstyle{\frac{1}{n-2}} (R_{ij}+R_{01} \delta_{ij})  + \textstyle{\frac{1}{(n-1)(n-2)}} R \delta_{ij} 
\, ,  \label{11i}\\
  %--------------------------------------------------------
\fl
D N_{ij} & = - N_{ik} L_{kj}  - C_{0j1i} 
%- {\textstyle{\frac{1}{n-2}}} (R_{ij}+R_{01}\delta_{ij})+  \textstyle{\frac{1}{(n-1)(n-2)}} R \delta_{ij} 
\, , \label{11j}\\
  %--------------------------------------------------------
\fl
\delta_{[j|} L_{i|k]}& =  L_{1[j|} L_{i|k]}  
+ L_{i1} L_{[jk]} +  L_{il} \Ml_{{[j}{k]} }
+ L_{l[j|}  \Ml_{{i|}{k]}}   
% - \textstyle{\frac{1}{2}} C_{0ijk} - \textstyle{\frac{1}{n-2}} R_{0[j} \delta_{k]i}
\, , \label{11k}\\ %\label{Lmmm}  \\
  %-------------------------------------------------------- 
\fl
\delta_{[j|} N_{i|k]} &=  - L_{1[j|} N_{i|k]}  
+ N_{i1} L_{[jk]}   +  N_{il} \Ml_{{[j}{k]}} 
%\nonumber\\
 % \qquad\qquad\qquad\qquad \!\!\!\!\!\!\!\! \!\!\!\!\!\!\ {}
+ N_{l[j|}  \Ml_{{i|}{k]}}   
 - \textstyle{\frac{1}{2}}C_{1ijk} %- \textstyle{\frac{1}{n-2}} R_{1[j} \delta_{k]i}
\, ,\label{11l} \\ 
 %--------------------------------------------------------
\fl
D \Mi_{{j}{1}}& =    
- \Mi_{{j}{k}} L_{k1}  - C_{01ij}\, ,\label{11m}  \\
  %--------------------------------------------------------
\fl
D \Mi_{{j}{k}} 
&=   - \Mi_{{j}{l}} L_{lk} 
%- C_{0kij} -\textstyle{\frac{1}{n-2}}{2R_{0[i}\delta_{j]k}}
\, ,   \label{11n}\\
   %--------------------------------------------------------
\fl
\bigtriangleup \Mi_{{j}{k}} - \delta_{k} \Mi_{{j}{1}} 
&=  2N_{j[1|} L_{i|k]}  
+2 L_{j[1|} N_{i|k]} 
+ \Mi_{{j}{1}} (L_{1k}-L_{k1})
+ 2  \Mi_{{l}{[1|}} \Ml_{{j|}{k]}}
 \nonumber\\ 
 %\! \! \! \! \! \! \! \! \! \! \! \!  \! \! \! \! \! \!  \! \! \! \! \! \! 
%\! \! \! \! \! \! \!
%\qquad\qquad\qquad\qquad\quad
 &{} - \Mi_{{j}{l}} (N_{lk} + \Ml_{{k}{1}}  )  - C_{1kij} 
%-\textstyle{\frac{1}{n-2}}{ 2R_{1[i}\delta_{j]k} }
\, ,   \label{11o}\\
    %--------------------------------------------------------
\fl
\delta_{[k|} \Mi_{{j|}{l]}} 
&=  N_{i[l|} L_{j|k]}  
+ L_{i[l|} N_{j|k]} +L_{[kl]} \Mi_{{j}{1}} 
\nonumber \\%\label{Mmmm}\\ 
&{}
+   \Mi_{{p}{[k|}} \Mp_{{j|}{l]}} + \Mi_{{j}{p}}  \Mp_{{[k}{l]}} 
  - \textstyle{\frac{1}{2}} C_{ijkl} 
%-\textstyle{\frac{1}{n-2}} (\delta_{i[k}  R_{l]j}-\delta_{j[k}R_{l]i})
 \, .\label{11p}
\EEA
%\endnumparts

%--------------------------------------------------------------------

\section{Bianchi equations}
\label{appbianchi}

We present here Bianchi identities projected onto a parallelly propagated null frame (\ref{normalization}) for vacuum spacetimes
with a geodetic multiple WAND. General form of these identities  can be found in Appendix B in 
\cite{Pravdaetal04}.

\BEA
%  prvni 1-index
\fl
 D C_{101i} -\delta_i C_{0101}= 
-C_{0101} L_{i1}
- C_{01is} L_{s1}- 2C_{101s} L_{si} 
-  C_{0i1s} L_{s1}     ,\label{B1}\\
%\EEA
%---------------------------
%2-index
%\BEA
\fl
%1. 2-index
- \T C_{01ij}+2\delta_{[j|}C_{101|i]} =
2C_{101[j|} L_{1|i]}  +  2C_{101[i|} L_{|j]1} +  2C_{1[i|1s} L_{s|j]}+  C_{1sij} L_{s1}  
\nonumber\\
\fl 
+  2C_{0101} N_{[ji]}
 +  2C_{01[i|s} N_{s|j]} +  2C_{0s1[j|} N_{s|i]}  
%\nonumber \\
%\fl &
  +  2C_{01[i|s} \Ms_{|j]1}    
 +  2C_{101s} \Ms_{[ji]}   ,\label{B2}\\
%\EEA
%-----
%\BEA
\fl
%2. 2-index
-D C_{01ij} =  2C_{0101} L_{[ij]} +  2C_{01[i|s} L_{s|j]}     
+2C_{0[i|1s} L_{s|j]}    ,\label{B3}\\
%\EEA
%--------
%\BEA
%3. 2-index
\fl
D C_{1i1j}-\T C_{0j1i}-\delta_j C_{101i}=
2C_{101i} L_{[1j]}+  2C_{1i[j|s} L_{s|1]}+  C_{0101} N_{ij}  \nonumber\\
\fl  
 -  C_{01is} N_{sj}+  C_{0s1i} N_{sj}   
+  C_{0j1s} \Ms_{i1}+ 2 C_{0s1i} \Ms_{j1}  +  C_{101s} \Ms_{ij}, \label{B4} \\    
%\EEA
%-----------
%\BEA
\fl
%4. 2-index
D C_{0i1j}=   
-C_{0101} L_{ij}  
-  C_{0i1s} L_{sj}-C_{01is} L_{sj}
  ,\label{B5}\\
%\EEA
%-------------------------------------------------------
%\BEA
\fl
%1. 3-index
-D C_{1kij}-\delta_k C_{01ij}=
  - C_{01ij} L_{k1}  +  2C_{0k1[i} L_{j]1}  +  2C_{101[i} L_{j]k} \nn\\
\fl
  + 2 C_{[1|sij} L_{s|k]}
 +  2C_{01[i|s} \Ms_{|j]k}  ,\label{B6}  \\
%\EEA
%-----------
%\BEA
\fl
%2. 3-index
2\delta_{[k|} C_{0i1|j]}= 2C_{0i1[j|} L_{|k]1}- C_{01jk} L_{i1}  +  2C_{101[j|} L_{i|k]}    
+  2C_{1[k|is} L_{s|j]}  \nn\\
\fl
- C_{isjk} L_{s1} 
+  2C_{0i1s} \Ms_{[kj]} +  2C_{0s1[k|} \Ms_{i|j]}   ,\label{B7}\\
%\EEA
%------------
%\BEA
\fl
%3. 3-index
0=0,\label{B8}\\
%\EEA
%-------
%\BEA
\fl
%4. 3-index
DC_{1ijk}+2\delta_{[k|}C_{0|j]1i}\!=\! 2C_{101i} L_{[jk]}      
+  2C_{1i[k|s} L_{s|j]} 
% \nn\\
%\fl
+  2C_{0[k|1s} \Ms_{i|j]}   - 2C_{0s1i} \Ms_{[jk]},\label{B9}\\
%\EEA
%-----------
%\BEA
\fl
%5. 3-index
\T C_{1ijk}+2\delta_{[k|}C_{1i1|j]}=
  2C_{1i1[j} L_{k]1}+  4C_{1i1[k|} L_{1|j]}- C_{1ijk} L_{11} +  C_{01jk} N_{i1} \nn\\
\fl
   +2C_{0[j|1i} N_{|k]1}  +  2C_{101[k|} N_{i|j]}  +  2C_{101i} N_{[kj]}  
+  2C_{1i[k|s} N_{s|j]}+  2C_{1[k|is} N_{s|j]}  \nn \\ \fl
- C_{isjk} N_{s1} 
- 2C_{1i1s} \Ms_{[jk]}  +  2C_{1[k|1s} \Ms_{i|j]}  +  2C_{1i[k|s} \Ms_{|j]1}  
- C_{1sjk} \Ms_{i1} ,\label{B10}\\
%\EEA
%--------
%\BEA
\fl
%6. 3-index
-\delta_{\{i|}C_{01|jk\}}=
C_{101\{i} L_{jk\}}  - C_{101\{i} L_{kj\}}  
+  C_{1s\{ij|} L_{s|k\}} \nn\\ \fl 
+  C_{01\{i|s} \Ms_{|jk\}} - C_{01\{i|s} \Ms_{|kj\}}  ,\label{B11}\\
%\EEA
%------------------------------------------------------
%
%\BEA
\fl
%1. 4-index
-DC_{ijkm}\! =\!
2C_{01ij} L_{[km]}  +  2C_{0[k|1j} L_{i|m]} +  2C_{0[m|1i} L_{j|k]} %\nn\\ \fl
%-C_{0m1j} L_{ik} 
+  2C_{ij[k|s} L_{s|m]} ,\label{B12}\\
%\EEA
%-------
%\BEA
\fl
%2. 4-index
-\T C_{ijkm}+2\delta_{[k|}C_{1|m]ij}=
2C_{1i[1|m} L_{j|k]} +  2C_{1[j|1k} L_{|i]m}  
+2C_{1j[k|m} L_{i|1]} +  2C_{1kij} L_{[1m]} \nn \\ \fl
  +  2C_{1mij} L_{[k1]}  
- 2C_{01ij} N_{[km]}   +  2C_{0[i|1m} N_{|j]k}
 +  2C_{0[j|1k} N_{|i]m}
  +  2C_{ij[k|s} N_{s|m]}\nn \\ \fl  
+\!  2C_{1k[i|s} \Ms_{|j]m} \!
 +\!  2C_{1m[j|s} \Ms_{|i]k} \! 
+ \! 2C_{1sij} \Ms_{[km]}\! +\!  2C_{ij[k|s} \Ms_{|m]1}\! +\!  2C_{[i|skm} \Ms_{|j]1}    ,\label{B13}\\
%\EEA
%
%\BEA
\fl
%3. 4-index
\delta_{\{j|}C_{1i|mk\}}=  
C_{1i1\{j} L_{mk\}} - C_{1i1\{j} L_{km\}} +  C_{1i\{jk|} L_{1|m\}} 
+  C_{01\{jm|} N_{i|k\}}   
- C_{0\{j|1i} N_{|km\}} \nn\\ \fl
 +  C_{0\{j|1i} N_{|mk\}} +  C_{is\{jk|} N_{s|m\}}  
- C_{1i\{j|s} \Ms_{|mk\}}\!
 +  C_{1i\{j|s} \Ms_{|km\}} \! 
+  C_{1s\{jk|} \Ms_{i|m\}} ,\label{B14}\\
%\EEA
%
%\BEA
\fl
%4. 4-index
0=
C_{01\{jk|} L_{i|m\}}
- C_{0i1\{j} L_{km\}}  +  C_{0i1\{j} L_{mk\}} 
+  C_{is\{jk|} L_{s|m\}} ,\label{B15}\\
%\EEA
%--------------------------------------------------
%\BEA
\fl
%5-index
\delta_{\{ k|}C_{ij|nm\}}=
C_{1j\{km|} L_{i|n\}}- C_{1i\{km|} L_{j|n\}} - C_{1\{k|ij} L_{|mn\}} +  C_{1\{k|ij} L_{|nm\}}  
\nn \\ \fl 
  +  C_{ij\{k|s} \Ms_{|mn\}} - C_{ij\{k|s} \Ms_{|nm\}}+  C_{is\{km|} \Ms_{j|n\}}- C_{js\{km|} \Ms_{i|n\}}  . \label{B16}
\EEA

%-----------------------------------------------------------------------

%---------------------------------------------
%-----------------------------------------------------------------------
\section{
The Ricci and Bianchi equations for five-dimensional type N spacetimes  with all $L_{1i}=0$}
\label{appN5nulL1i}

Ricci equations (\ref{11d}) 
\BEA
n_{32}^0&=&n_{23}^0,\label{Ricci_d_1s}\\
n_{24}^0&=&0,\label{Ricci_d_5s}\\
n_{34}^0&=&0,\label{Ricci_d_6s}
\EEA
Ricci equations (\ref{11i}) 
\BEA
{ U}^0&=&n_{22}^0=n_{33}^0,\label{Ricci_i_Us}\\
n_{23}^0&=&0=n_{32}^0,\\
0&=&n_{24}^0+\mDo{41} ,\label{Ricci_i_3s}\\
0&=&n_{34}^0+\mTo{41} .\label{Ricci_i_4s}
\EEA
Ricci equations (\ref{11k}) 
\BEA
{   \omega}_2^0&=&0={   \omega}_3^0,\label{Ricci_k_1s}\\
{   \omega}_4^0&=&\mCo{22} =\mCo{33} ,\label{Ricci_k_2s}\\
\mDo{43} &=&0=\mTo{42} ,\label{Ricci_k_3s}\\
\mDo{44} &=&0=\mTo{44} .\label{Ricci_k_4s}
\EEA
Ricci equations (\ref{11c})
\BEA
{\xi^{\alpha 0}_{2}}    l_{11}^0,_\alpha&=& 
-   n_{21}^0,\label{Ricci_c_l11,2s}\\
{\xi^{\alpha 0}_{3}}    l_{11}^0,_\alpha&=& 
-   n_{31}^0,\label{Ricci_c_l11,3s}\\
{\xi^{\alpha 0}_{4}}    l_{11}^0,_\alpha
&=& 0.
\label{Ricci_c_l11,4s}
\EEA
Ricci equations (\ref{11l}) 
with $n_{24}=n_{34}=0$ from (\ref{Ricci_d_5s}), (\ref{Ricci_d_6s}), with (\ref{m234_0})
\BEA
\ \  \quad\quad\quad\quad
{\xi^{\alpha 0}_{2}}  n_{22}^{0},_\alpha &=&n_{42}^0\mCo{22} ,\label{Ricci_l_1s}\\
\ \ 
 \quad\quad\quad\quad
{\xi^{\alpha 0}_{3}}  n_{22}^{0},_\alpha &=&n_{43}^0\mCo{22} ,\label{Ricci_l_2s}\\
\ \ 
\quad \quad\quad\quad
{\xi^{\alpha 0}_{4}}  n_{22}^{0},_\alpha &=&
n_{44}^0\mCo{22} ,\label{Ricci_l_3s}\\
{\xi^{\alpha 0}_{3}}  n_{42}^{0},_\alpha -{\xi^{\alpha 0}_{2}}  n_{43}^{0},_\alpha &=&
n_{42}^0\mDo{32} -n_{43}^0\mTo{23} ,\label{Ricci_l_4s}\\
{\xi^{\alpha 0}_{2}}  n_{44}^{0},_\alpha -{\xi^{\alpha 0}_{4}}  n_{42}^{0},_\alpha &=&
0,\label{Ricci_l_5s}\\
{\xi^{\alpha 0}_{3}}  n_{44}^{0},_\alpha -{\xi^{\alpha 0}_{4}}  n_{43}^{0},_\alpha&=&
0.\label{Ricci_l_10s}
\EEA
Ricci equations (\ref{11h}),  
with $n_{24}=n_{34}=0$, 
\BEA
\fl
n_{22}^0,_u+   X^{\alpha 0}n_{22}^0,_\alpha - {\xi^{\alpha 0}_{2}}    n_{21}^{0},_\alpha &=&-2   l_{11}^0n_{22}^0
+    n_{31}^{0}\mTo{22} 
-p^0,\label{Ricci_h_n22u_n21S}\\
\fl
\quad \quad\quad\quad\quad\quad\quad\quad   {\xi^{\alpha 0}_{3}}    n_{21}^{0},_\alpha &=&
    n_{31}^{0}\mDo{33} +\Pi^0
,\label{Ricci_h_2S}\\
 \fl
 \quad \quad\quad\quad\quad\quad\quad\quad
 {\xi^{\alpha 0}_{4}}    n_{21}^{0},_\alpha &=&
 0,\label{Ricci_h_3S}\\ 
\fl
\quad \quad\quad\quad\quad\quad\quad\quad
{\xi^{\alpha 0}_{2}}    n_{31}^{0},_\alpha &=&
    n_{21}^{0}\mTo{22} +\Pi^0
,\label{Ricci_h_4S}\\
\fl
n_{22}^0,_u+   X^{\alpha 0}n_{22}^0,_\alpha - {\xi^{\alpha 0}_{3}}    n_{31}^{0},_\alpha &=&-2   l_{11}^0n_{22}^0
+   n_{41}^{0}
\mCo{22} 
+    n_{21}^{0}\mDo{33}
+p^0,\label{Ricci_h_n22u_n31S}\\
\fl
\quad \quad\quad\quad\quad\quad\quad\quad
{\xi^{\alpha 0}_{4}}    n_{31}^{0},_\alpha &=&
0.\label{Ricci_h_6S} 
\EEA
Ricci equations (\ref{11p}) 
\BEA
\fl
\quad \quad\quad\quad\quad\quad
{\xi^{\alpha 0}_{4}}  \mDo{32} ,_\alpha &=&
0,\label{Ricci_p_1s}\\
\fl
\quad \quad\quad\quad\quad\quad
{\xi^{\alpha 0}_{4}}  \mDo{33} ,_\alpha &=&
0,\label{Ricci_p_2s}\\
\fl
{\xi^{\alpha 0}_{3}}  \mDo{32} ,_\alpha -{\xi^{\alpha 0}_{2}}  \mDo{33} ,_\alpha &=&
2n_{22}^0+({\mCo{22} })^2+({\mDo{32} })^2+({\mDo{33} })^2,
\label{Ricci_p_3s}\\
\fl
\quad \quad\quad\quad\quad\quad
{\xi^{\alpha 0}_{2}}  \mCo{22} ,_\alpha &=&-n_{42}^0,\label{Ricci_p_m422_2s}\\
\fl
\quad \quad\quad\quad\quad\quad
{\xi^{\alpha 0}_{3}}  \mCo{22} ,_\alpha &=&-n_{43}^0,\label{Ricci_p_m422_3s}\\
\fl
\quad \quad\quad\quad\quad\quad
{\xi^{\alpha 0}_{4}}  \mCo{22} ,_\alpha &=&-n_{44}^0
.\label{Ricci_p_m422_4s}
\EEA
Ricci equations (\ref{11o})
\BEA
\fl
\mDo{32} ,_u+   X^{\alpha 0}\mDo{32} ,_\alpha - {\xi^{\alpha 0}_{2}} \mDo{31} ,_\alpha &=&
-\mDo{32}  
   l_{11}^0
+   n_{31}^0
+\mDo{33} \mDo{31} ,\label{Ricci_o_1s}\\
\fl
\mDo{33} ,_u+   X^{\alpha 0}\mDo{33} ,_\alpha - {\xi^{\alpha 0}_{3}} \mDo{31} ,_\alpha &=&
-\mDo{33}  
   l_{11}^0
-   n_{21}^0
-\mDo{32} \mDo{31} ,\label{Ricci_o_2s}\\
\fl
\quad \quad\quad\quad\quad\quad\quad\quad \quad\quad
{\xi^{\alpha 0}_{4}} \mDo{31} ,_\alpha &=&
0,\label{Ricci_o_3s}\\
\fl
\quad \quad\quad\quad\quad\quad
\mDo{42} ,_u+   X^{\alpha 0}\mDo{42} ,_\alpha
&=&
-\mDo{42}  
   l_{11}^0
+   n_{41}^0
.\label{Ricci_o_4s}
\EEA

Commutators (\ref{com_x_delxi}) 
\BEA
{\xi^{\beta 0}_{2}} {\xi^{\alpha 0}_{3}},_\beta- {\xi^{\beta 0}_{3}} {\xi^{\alpha 0}_{2}},_\beta&=&
\mTo{22}  {\xi^{\alpha 0}_{2}}-\mDo{33}  {\xi^{\alpha 0}_{3}},\label{com_x_delxi_1s}\\
{\xi^{\beta 0}_{2}} {\xi^{\alpha 0}_{4}},_\beta - {\xi^{\beta 0}_{4}} {\xi^{\alpha 0}_{2}},_\beta &=&
{\xi^{\alpha 0}_{2}}
\mCo{22} ,\label{com_x_delxi_2s}\\
{\xi^{\beta 0}_{3}} {\xi^{\alpha 0}_{4}},_\beta - {\xi^{\beta 0}_{4}} {\xi^{\alpha 0}_{3}},_\beta &=&
{\xi^{\alpha 0}_{3}}
\mCo{22} ,\label{com_x_delxi_3s}
\EEA
commutators (\ref{com_x_delX}) 
\BEA
-{\xi^{\alpha 0}_{2}},_u- X^{\beta 0}{\xi^{\alpha 0}_{2}},_\beta + {\xi^{\beta 0}_{2}}  X^{\alpha 0},_\beta
&=& {\xi^{\alpha 0}_{2}}   l_{11}^0
+n_{42}^0 {\xi^{\alpha 0}_{4}}-\mDo{31}  {\xi^{\alpha 0}_{3}}
,\label{com_x_delX_1s}\\
-{\xi^{\alpha 0}_{3}},_u- X^{\beta 0}{\xi^{\alpha 0}_{3}},_\beta + {\xi^{\beta 0}_{3}}  X^{\alpha 0},_\beta
&=& {\xi^{\alpha 0}_{3}} l_{11}^0
+n_{43}^0 {\xi^{\alpha 0}_{4}}+\mDo{31}  {\xi^{\alpha 0}_{2}}
,\label{com_x_delX_2s}\\
-{\xi^{\alpha 0}_{4}},_u- X^{\beta 0}{\xi^{\alpha 0}_{4}},_\beta + {\xi^{\beta 0}_{4}} X^{\alpha 0},_\beta &=& 
{\xi^{\alpha 0}_{4}}
n_{44}^0, \label{com_x_delX_3s}
\EEA
commutators 
(\ref{com_r_delom}) 
\BEA
{\xi^{\alpha 0}_{2}} \omega_4^0,_\alpha &=&-n_{42}^0,\label{com_r_delom_1s}\\
{\xi^{\alpha 0}_{3}} \omega_4^0,_\alpha &=&-n_{43}^0,\label{com_r_delom_2s}
\EEA
commutators
(\ref{com_r_delU}) 
\BEA
\quad \quad\quad\quad\quad\quad\quad\quad
{\xi^{\alpha 0}_{2}}  U^0,_\alpha &=&n_{42}^0\omega_4^0,\label{com_r_delU_1s}\\
\quad \quad\quad\quad\quad\quad\quad\quad
{\xi^{\alpha 0}_{3}}  U^0,_\alpha &=&n_{43}^0\omega_4^0,\label{com_r_delU_2s}\\
-\omega_4^0,_u- X^{\alpha 0}\omega_4^0,_\alpha +  {\xi^{\alpha 0}_{4}}  U^0,_\alpha &=&n_{44}^0\omega_4^0
+\mCo{22} 
 l_{11}^0
+ n_{41}^0.
\label{com_r_delU_3s}
\EEA

Bianchi equations (\ref{B10}) %= (24) in \cite{Pravdaetal04}) 
\BEA
-{\xi^{\alpha 0}_{2}}  p^0,_\alpha -{\xi^{\alpha 0}_{3}}  \Pi^0,_\alpha &=&2p^0\mDo{33}
+2\Pi^0\mTo{22} ,\label{Bianchi_B10_1S}\\\
\ {\xi^{\alpha 0}_{3}}  p^0,_\alpha -{\xi^{\alpha 0}_{2}}  \Pi^0,_\alpha &=&2p^0\mDo{32} 
+2\Pi^0\mDo{33} ,\label{Bianchi_B10_2S}\\
\quad\quad\quad\quad\quad
{\xi^{\alpha 0}_{4}}  p^0,_\alpha &=&2\Pi^0\mDo{34}  
,\label{Bianchi_B10_3S}\\
\quad\quad\quad\quad\quad
{\xi^{\alpha 0}_{4}}  \Pi^0,_\alpha &=&-p^0 \mDo{34} 
,\label{Bianchi_B10_3bS}\\
p^0\mCo{23} =p^0\mCo{32} &=& 0,\\
\Pi^0 \mCo{32} = \Pi^0 \mCo{23} &=&0 ,\label{Bianchi_B10_4s}\\
p^0\mDo{44} +\Pi^0\mTo{44} &=& -p^0\mTo{44} +\Pi^0\mDo{44} =0.
\EEA

\section*{References}

%\bibliographystyle{JHEP}
%\bibliographystyle{unsrt}
%\bibliography{bibl,Vojta}

%\end{document}

%--------------------------------------------------------------------------

%\bibliographystyle{unsrt}
%\bibliography{bibl,Vojta}

\end{document}